\documentclass[twocolumn]{aastex631}

\shorttitle{A Headwind-like Helium Outflow from TOI-3884b}
\shortauthors{J. P. Ninan et al.}
\newcommand{\unit}[1]{\ensuremath{\, \mathrm{#1}}}

\begin{document}

\title{A Headwind-like Helium Outflow from TOI-3884b}

\newcommand{\PSUAA}{Department of Astronomy and Astrophysics, 525 Davey Laboratory, 251 Pollock Road, Penn State, University Park, PA, 16802, USA}
\newcommand{\PSUCEHW}{Center for Exoplanets and Habitable Worlds, 525 Davey Laboratory, 251 Pollock Road, Penn State, University Park, PA, 16802, USA}
\newcommand{\PSUARC}{Astrobiology Research Center, 525 Davey Laboratory, 251 Pollock Road, Penn State, University Park, PA, 16802, USA}
\newcommand{\PSETI}{Penn State Extraterrestrial Intelligence Center, 525 Davey Laboratory, 251 Pollock Road, Penn State, University Park, PA, 16802, USA}
\newcommand{\UA}{Steward Observatory, University of Arizona, 933 N.\ Cherry Ave, Tucson, AZ 85721, USA}
\newcommand{\UAA}{Department of Astronomy and Steward Observatory, University of Arizona, Tucson, AZ 85721, USA}
\newcommand{\Penn}{Department of Physics and Astronomy, University of Pennsylvania, 209 S 33rd St, Philadelphia, PA 19104, USA}
\newcommand{\Caltech}{Department of Astronomy, California Institute of Technology, 1200 E California Blvd, Pasadena, CA 91125, USA}
\newcommand{\STScI}{Space Telescope Science Institute, 3700 San Martin Dr, Baltimore, MD 21218, USA}
\newcommand{\JHU}{Department of Physics and Astronomy, Johns Hopkins University, 3400 N Charles St, Baltimore, MD 21218, USA}
\newcommand{\GoddardESAL}{Exoplanets and Stellar Astrophysics Laboratory, NASA Goddard Space Flight Center, Greenbelt, MD 20771, USA}
\newcommand{\GoddardISTD}{Instrument Systems and Technology Division, NASA Goddard Space Flight Center, Greenbelt, MD 20771, USA}
\newcommand{\GSFC}{NASA Goddard Space Flight Center, Greenbelt, MD 20771, USA}
\newcommand{\NOAO}{U.S. National Science Foundation National Optical-Infrared Astronomy Research Laboratory, 950 N.\ Cherry Ave., Tucson, AZ 85719, USA}
\newcommand{\UW}{Wisconsin address goes here}
\newcommand{\Macquarie}{School of Mathematical and Physical Sciences, Macquarie University, Balaclava Road, North Ryde, NSW 2109, Australia}
\newcommand{\MacquarieCentre}{Astrophysics and Space Technologies Research Centre, Macquarie University, Balaclava Road, North Ryde, NSW 2109, Australia}
\newcommand{\NIST}{National Institute of Standards \& Technology, 325 Broadway, Boulder, CO 80305, USA}
\newcommand{\CUBoulderEE}{Electrical, Computer \& Energy Engineering, 440 UCB, University of Colorado, Boulder, CO 80309, USA}
\newcommand{\CUBoulder}{Department of Physics, 390 UCB, University of Colorado, Boulder, CO 80309, USA}
\newcommand{\JPL}{Jet Propulsion Laboratory, California Institute of Technology, 4800 Oak Grove Drive, Pasadena, California 91109}
\newcommand{\MIT}{Kavli Institute for Astrophysics and Space Research, Massachusetts Institute of Technology, Cambridge, MA, USA}
\newcommand{\UCI}{Department of Physics \& Astronomy, The University of California, Irvine, Irvine, CA 92697, USA}
\newcommand{\Carleton}{Carleton College, One North College St., Northfield, MN 55057, USA}
\newcommand{\Carnegie}{Earth and Planets Laboratory, Carnegie Science, 5241 Broad Branch Road, NW, Washington, DC 20015, USA}
\newcommand{\PSUICS}{Institute for Computational and Data Sciences, Penn State, University Park, PA, 16802, USA}
\newcommand{\PSUCASt}{Center for Astrostatistics, 525 Davey Laboratory, 251 Pollock Road, Penn State, University Park, PA, 16802, USA}
\newcommand{\NESSF}{NASA Earth and Space Science Fellow}
\newcommand{\Princeton}{Department of Astrophysical Sciences, Princeton University, 4 Ivy Lane, Princeton, NJ 08540, USA}
\newcommand{\RUSSELL}{Henry Norris Russell Fellow}
\newcommand{\IAS}{Institute for Advance Study, 1 Einstein Drive, Princeton, NJ 08540, USA}
\newcommand{\Tsinghua}{Department of Astronomy, Tsinghua University, Beijing 100084, China}
\newcommand{\FlatironCCA}{Center for Computational Astrophysics, Flatiron Institute, 162 Fifth Avenue, New York, NY 10010, USA}
\newcommand{\ETH}{ETH Zurich, Institute for Particle Physics \& Astrophysics, Zurich, Switzerland}
\newcommand{\TIFR}{Department of Astronomy and Astrophysics, Tata Institute of Fundamental Research, Homi Bhabha Road, Colaba, Mumbai 400005, India}
\newcommand{\UCLA}{Department of Physics \& Astronomy, University of California Los Angeles, Los Angeles, CA 90095, USA}
\newcommand{\UAm}{Anton Pannekoek Institute for Astronomy, 904 Science Park, University of Amsterdam, Amsterdam, 1098 XH}
\newcommand{\UMd}{Department of Physics \& Astronomy, 1023 University Drive, University of Minnesota Duluth, Duluth, MN 55812, USA}
\newcommand{\UIUC}{Department of Astronomy, University of Illinois at Urbana-Champaign, Urbana, IL 61801, USA}
\newcommand{\Amherst}{Department of Physics and Astronomy, Amherst College, 25 East Drive, Amherst, MA 01002, USA}
\newcommand{\UTAustin}{Department of Astronomy, The University of Texas at Austin, 2515 Speedway, Austin, TX 78712, USA}
\newcommand{\DTU}{DTU Space, Elektrovej, Building 328, DK-2800 Kgs. Lyngby, Denmark}
\newcommand{\Schmidt}{Astrophysics \& Space Center, Schmidt Sciences, New York, NY 10011, USA}


\correspondingauthor{Joe P.\ Ninan}
\email{joe.ninan@tifr.res.in}

\author[0000-0001-8720-5612]{Joe P.\ Ninan}
\affil{Department of Astronomy and Astrophysics, Tata Institute of Fundamental Research, Homi Bhabha Road, Colaba, Mumbai 400005, India}

\author[0000-0002-2990-7613]{Jessica Libby-Roberts}
\affiliation{\PSUAA}
\affiliation{\PSUCEHW}
\affiliation{Department of Physics and Astronomy, University of Tampa, Tampa, FL 33606}

\author[0000-0001-9596-7983]{Suvrath Mahadevan}
\affiliation{\PSUAA}
\affiliation{\PSUCEHW}

\author[0000-0003-3061-4591]{Oleg Kochukhov}
\affil{Department of Physics and AstronomyUppsala University, Box 516, Uppsala,  751 20, Sweden}

\author[0000-0001-9626-0613]{Daniel M.\ Krolikowski}
\affiliation{\UA}

\author[0000-0001-7409-5688]{Gudmundur Stefansson}
\affiliation{\Schmidt}
\affiliation{Anton Pannekoek Institute for Astronomy, University of Amsterdam, Amsterdam, 1098 XH, The Netherlands}

\author[0000-0001-8401-4300]{Shubham Kanodia}
\affiliation{\Carnegie}

\author[0000-0003-2066-8959]{Jaume Orell-Miquel}
\affil{Department of Astronomy, University of Texas at Austin, 2515 Speedway, Austin, TX 78712, USA}

\author[0000-0002-4404-0456]{Caroline V. Morley}
\affil{Department of Astronomy, University of Texas at Austin, 2515 Speedway, Austin, TX 78712, USA}


\author[0000-0002-1002-3674]{Kevin France} 
\affil{Laboratory for Atmospheric and Space Physics, University of Colorado, 600 UCB, Boulder, CO 80309, USA}

\author[0000-0002-7119-2543]{Girish Duvvuri}
\affil{Department of Astrophysical and Planetary Science, University of Colorado, 389 UCB, Boulder, CO 80309, USA}

\author[0000-0003-4384-7220]{Chad F.\ Bender}
\affiliation{\UA}

\author[0000-0001-9662-3496]{William D. Cochran}
\affil{McDonald Observatory and Department of Astronomy, The University of Texas at Austin, USA}
\affil{Center for Planetary Systems Habitability, The University of Texas at Austin, USA}

\author[0000-0003-1439-2781]{Megan Delamer}
\affiliation{\PSUAA}
\affiliation{\PSUCEHW}

\author[0000-0002-2144-0764]{Scott A.\ Diddams}
\affiliation{\CUBoulderEE}
\affiliation{\CUBoulder}

\author[0000-0003-1312-9391]{Samuel Halverson}
\affiliation{\JPL}

\author[0000-0002-1664-3102]{Fred Hearty}
\affiliation{\PSUAA}
\affiliation{\PSUCEHW}

\author[0000-0003-1263-8637]{Leslie Hebb}
\affil{Physics Department, Hobart and William Smith Colleges, 300 Pulteney Street, Geneva, NY 14456, USA}
\affil{Department of Astronomy, Cornell University, 245 East Ave, Ithaca, NY 14850, USA}

\author[0000-0002-0048-2586]{Andrew Monson}
\affiliation{\UA}

\author[0000-0003-0149-9678]{Paul Robertson}
\affiliation{\UCI}

\author[0000-0001-8127-5775]{Arpita Roy}
\affiliation{\Schmidt}

\author[0000-0002-4046-987X]{Christian Schwab}
\affiliation{\Macquarie}



\begin{abstract}

Constraining atmospheric loss is crucial for understanding the evolution of planetary atmospheres. We report the detection of an evaporating exosphere from TOI-3884b, a super-Neptune in a polar orbit around an M4 dwarf star. We infer this evaporation from a 6.8$\pm$0.27\% excess transit depth in the He I 10830 \AA\, line, observed during five transits of TOI-3884b with the Habitable Zone Planet Finder spectrometer. While we observe outflow signatures starting as early as 5 hours before the onset of the planet's optical transit,  no significant post-transit tail is detected. Hitherto, such a pre-transit headwind without a significant post-transit tail has only been detected from HAT-P-67b, a hot Saturn orbiting an F-type star. This asymmetric outflow morphology supports irradiation-driven mass loss over core-heated mass loss, and we explore whether this could be caused by sub-Alfvén star-planet interaction. The detection of such complex outflow morphologies is now opening a new window, enabling us to test theories of star-planet interactions and planetary magnetic fields with observations.

\end{abstract}

\keywords{Exoplanet astronomy (486), Exoplanet atmospheric evolution (2308), Exosphere (499), Star-planet interactions (2177), High resolution spectroscopy (2096)}


\section{Introduction} \label{sec:intro}

Atmospheric escape drives the evolution and composition of planetary atmospheres, from primordial atmospheres to the final atmospheric composition of planets that we observe. Direct measurement of this phenomenon is crucial for understanding how planetary atmospheres evolve, shaping the observed properties of planets and their demographics. To measure the extended but tenuous exospheres of planets via transmission spectroscopy, one needs to select absorption lines with the largest cross sections. Since neutral hydrogen is abundant in giant planetary atmospheres, Ly$\alpha$ provides the strongest absorption signal. \citet{vidalmadjar03} detected exospheric outflow from HD 209458b via excess absorption in the Ly$\alpha$ wings. A few other planetary exospheres have also been detected using Ly$\alpha$ 
\citep{Lecavelier10,kulow14,ehrenreich15,bourrier18,dosSantos19,Loyd25}. While Ly$\alpha$ is the strongest absorber, it suffers from the drawback that the interstellar medium is also full of neutral hydrogen. This, combined with geocoronal contamination, completely absorbs the zero-velocity component of Ly$\alpha$, leaving only the high-velocity wings detectable. \cite{seager00,turner16,oklopcic18} identified the He I 10830\AA\, absorption line of metastable helium as a potential alternative. If the conditions allow the formation of metastable helium atoms, the absorption line formed by the resonant scattering of 10830\AA\, photon is ideal, because it is not affected by the interstellar medium and has minimal contamination from the Earth's atmosphere. This implies that ground-based, high-resolution spectrographs on large telescopes can detect this signature. Initial measurements of the He I 10830\AA\, signature were made around F-,G-, and K- type star planets \citep{spake18,allart19,salz18,mansfield18, allart18,nortmann18,floriano19}. For these spectral types, the stellar XUV flux is sufficient to ionize helium and to form metastable helium atoms during the recombination cascade through triplet states. While the XUV flux from a late-type M-dwarf photosphere is not sufficient to ionize helium, the active chromospheres of some M-dwarfs produce enough XUV flux. The first detection of a planetary helium exosphere around an M-dwarf (GJ 3470b) was reported by \citet{Ninan20, Palle20}. These detections enabled measurements of planetary outflow rates around stars beyond the traditional FGK-type hosts.

The Habitable-zone Planet Finder (HPF) has a long-running campaign to detect these exospheres \citep{Ninan20,Zhang23,Santiago24,Orell26}.
TOI-3884b is a super-Neptune in a polar orbit around an active M4 dwarf with persistent spots \citep{Almenara22,Roberts23}. Since it has a low bulk density of 0.67 g cm$^{-3}$, it is a good  candidate for atmospheric mass loss. The transits of TOI-3884 show persistent starspot crossings, and several studies have confirmed that this spot complex lies near the stellar pole \citep{Roberts23,Tamburo25,Mori25,Chakraborty25}, and is partially occulted by the planet in the first half of its transit. Being fully convective and a low-mass star, its photosphere is not hot enough to produce sufficient XUV radiation to ionize helium. However, XUV radiation from active regions and the chromosphere could potentially produce enough metastable helium atoms to enable detection of the outflow. Irrespective of whether the outflow is core-powered or photoevaporation-driven, measurement of the outflow morphology can help constrain the mass-loss mechanisms in these late M-dwarf systems.
In Section \ref{sec:observations} we outline our in- and out-of-transit observations of TOI-3884b and the data reduction. Section \ref{sec:discussion} discusses the detection and modelling of the observed outflow rate. Finally, in Section \ref{sec:conclusion} we summarize the main results and provide insights into what is driving the outflow in these late M-dwarf systems.

\section{Observations and Data Reduction} \label{sec:observations}
We observed TOI-3884 with the Habitable Zone Planet Finder (HPF) spectrograph on the 10-m Hobby-Eberly Telescope (HET) at McDonald Observatory, USA. HPF is an ultra-stabilized extreme precision radial velocity spectrograph, operating in near-infrared (8079-12786 \AA) with a resolving power of $R \sim $55,000 \citep{mah2014SPIE,metcalf19,stefansson16}. HET has a fixed-altitude primary mirror and a movable prime-focus unit that tracks the star. This limits the observable altitude range of HET to 55 $\pm$ 8 degrees \citep{HETRamsey, HETqueue}. For TOI-3884 at declination +12.5 degrees, this implies a continuous maximum track length of 4080 seconds. We observed TOI-3884 across multiple nights spread over 4 years. We obtained 49 in-transit spectra across 5 transits, and 64 out-of-transit spectra randomly spread across the out-of-transit phase. Table \ref{tab:observationlog} lists our in-transit and out-of-transit observations of TOI-3884b.

\begin{table*}[t!]
\renewcommand{\thetable}{\arabic{table}}
\centering
\caption{HPF Observation Log of TOI-3884b used in this work} \label{tab:observationlog}
\begin{tabular}{cccl}
\tablewidth{0pt}
\hline
\hline
Observation JD & Exposure Time (s) & SNR & In/Out Tranist\\
\hline
\decimals
2459550.990452&969.1&64&Out\\
2459551.001865&969.1&73&Out\\
2459564.962886&969.1&60&Out\\
2459564.973614&969.1&65&Out\\
2459569.946142&969.1&54&Out\\
2459569.957349&969.1&60&Out\\
2459570.940155&969.1&52&Out\\
2459570.951242&969.1&37&Out\\
2459571.942175&969.1&63&Out\\
2459571.952683&969.1&61&Out\\
2459575.932589&969.1&37&Out\\
2459575.943844&969.1&65&Out\\
2459596.868282&969.1&49&Out\\
2459596.88009&969.1&69&Out\\
2459597.877765&969.1&72&Out\\
2459597.889182&969.1&72&Out\\
2459601.868763&969.1&60&In\\
2459601.88035&969.1&51&In\\
...&...&...&...\\
\hline
\end{tabular}
\tablecomments{This table is available in its entirety in machine-readable form online.}
\end{table*}

We optimized our scheduling to preferentially observe when the observer barycentric velocity shifts the telluric OH emission line and the water absorption line near 10833\AA\, away from the stellar and planetary rest frame position of He I 10830\AA\,.

The observed data were reduced using the standard PSU HPF pipeline. Briefly, the steps include conversion from raw up-the-ramp data to slope images (after bias-noise, and nonlinearity correction) using \texttt{HxRGProc} \citep{ninan18}. The slope image contains the photoelectrons collected in each pixel per second, as well as an estimated uncertainty for each pixel value. This error includes both detector and photon noise, and it is propagated through all the subsequent steps of the pipeline. The spectra are extracted using the QTH lamp continuum spectrum in the flat-relative optimal extraction algorithm \citep{Zechmeister14}. The extracted spectra are wavelength calibrated using Laser Frequency Comb (LFC) spectra obtained throughout the night. This wavelength calibration procedure is detailed in \citet{stefansson20}. The telluric absorption in the spectra were corrected using a grid of telluric models generated with LBLRTM \citep{Clough2005} interpolated to each frame's best fit precipitable water vapor column and convolved with an accurate representation of HPF's instrumental line spread function (Krolikowski et al. in prep.). The simultaneous sky spectrum from the sky fiber is interpolated onto a common wavelength grid and subtracted from the stellar science spectra. This subtraction is not perfect and leaves some residuals in the data. The OH sky-emission line strengths are correlated across the sky spectrum. Hence, we could remove the residuals from sky-subtraction, without overfitting the OH lines near 10830 \AA\, by modelling all the residuals of the sky-emission subtraction across the echelle order containing the 10830 \AA\,. We first created a stellar template spectrum by averaging all the spectra in barycentric velocity frame. This stellar spectrum was shifted to the telluric frame and subtracted from each epoch to obtain the residuals from the sky-subtraction. We performed a Principal Component Analysis (PCA) to obtain eigenvectors that represent the correlated residual structures in the OH sky-emission lines across the echelle order. Based on the PCA explained variance ratio, we selected six principal components. These vectors were linearly fit to each spectrum to remove residuals from the OH sky-emission-line subtraction across the echelle order. In addition, in the observations taken just before sunrise (after 3:30 am CST, 9:30 am UTC), we found an excess He I 10830\AA\, emission signature in the telluric rest frame. This emission originates from the Earth's exosphere \citep{Shefov64}. We inflated the error bars in this narrow emission-line region in Earth's rest frame for these epochs to render any residual signal from this feature statistically insignificant. 

To generate an out-of-transit template spectrum, we used all spectra taken outside a 7-hour window around the center of transit. This conservative out-of-transit window was chosen to avoid any potential helium escape beyond the white light transit duration of TOI-3884b (1.58 hours). All out-of-transit spectra were averaged together, weighted by the variance, to obtain a high-S/N out-of-transit template spectrum. This template spectrum was then used to divide each spectrum, yielding the ratio of in-transit to out-of-transit, i.e., the transmission spectra. A second-order polynomial across the echelle order containing 10830\AA\, was used to normalize epoch-to-epoch variations in the continuum that could arise from instrumental effects and changing sky conditions.

\begin{figure}
\begin{center}
\includegraphics[width=1.0\columnwidth]{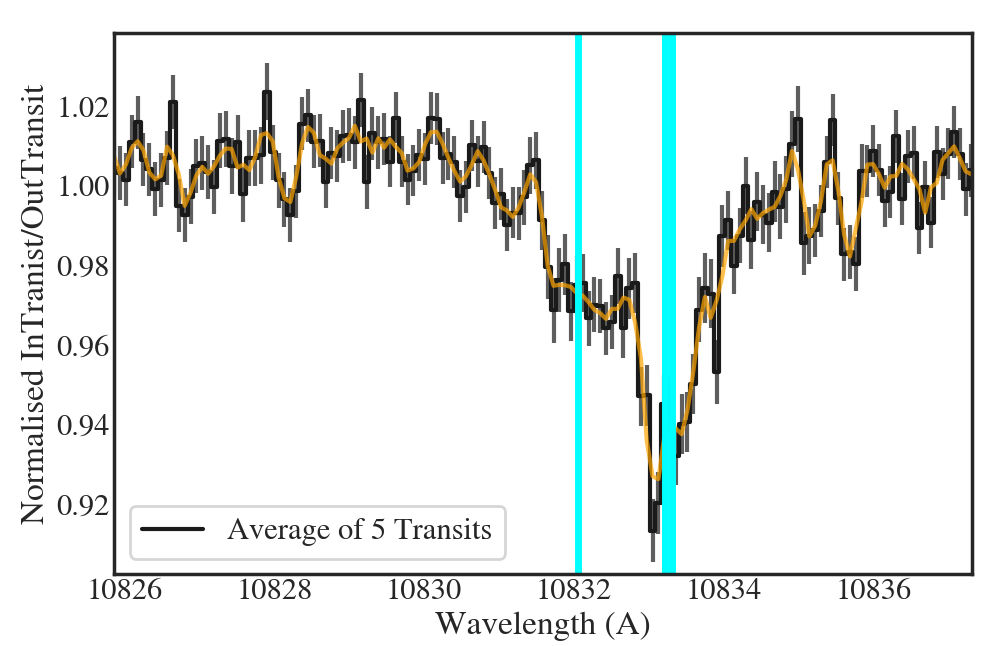}
\caption{This figure shows the average He I 10830\AA\, absorption profile from TOI-3884b during transit. 
The black curve is the weighted-average profile from all the five transit epochs. The orange curve is a 3-pixel-smoothed version of the same profile. The vacuum  rest wavelengths of the He I 10830\AA\, triplet lines are marked by the vertical cyan lines. The x-axis shows vacuum wavelength in the planet's rest frame.}
\label{fig:AvgHe10830Profile}
\end{center}
\end{figure}

Since our in-transit spectra were observed across multiple transits under different sky conditions, barycentric velocity, and S/N ratios, we used the following procedure to measure the excess transit depth for each epoch. First, all in-transit spectra were averaged together in planet's reference frame, weighted by the variance, to obtain the in-transit/out-of-transit He I 10830\AA\, transmission signature. This is shown in Figure \ref{fig:AvgHe10830Profile}. This average profile was then fit, using a weighted least-squares method (with pixel-by-pixel uncertainties) on individual epoch spectrum, to obtain the peak transit depth at each epoch. Since the shape of the absorption profile is fixed, the area inside the curve (i.e., equivalent width) also scales proportional to the transit depth. This procedure provides a robust and efficient estimate of the transit depth and corresponding equivalent widths for heteroskedastic transmission spectra. One major drawback of this technique is that it assumes the absorption profile shape is not variable and is the same across all transit epochs. Within the uncertainties of our single-epoch-spectra we do not see any statistically significant variations.
The TOI-3884b orbital and planetary parameters used in this work are summarized in Table \ref{table:TOI3884properties}.

\begin{table*}[t!]
\renewcommand{\thetable}{\arabic{table}}
\centering
\caption{Orbital parameters of TOI-3884b used in calculation} \label{table:TOI3884properties}
\begin{tabular}{ccll}
\tablewidth{0pt}
\hline
\hline
Parameter & Value & Description & Reference\\
\hline
\decimals
$\gamma$            & 8.3444 $\pm$ 0.0015                     & Absolute stellar RV ($\unit{km\:s^{-1}}$) & \citet{Almenara22}  \\
$M_*$               & 0.298 $\pm$0.018              & Stellar mass (M$_\odot$)                  & \citet{Roberts23}   \\
$T_0$               & 2459556.51661753 $\pm$0.00025 & Transit midpoint ($\mathrm{BJD_{TDB}}$)   & \citet{Roberts23} \\
$T_{14}$            & 0.06609 $\pm$0.002          & Transit duration (days)                   & \citet{Roberts23} \\
$P$                 & 4.54458889 $\pm$0.0000098    & Period (days)                             & \citet{Roberts23} \\
$e$                 & 0.06 $\pm$0.05            & Eccentricity                              & \citet{Roberts23} \\
$\omega$            & -1.95$^{+4.28}_{-0.04}$      & Argument of periastron (radians)          & \citet{Roberts23} \\
$K$                 & 28.03$^{+6.06}_{-6.23}$      & RV semi amplitude ($\unit{m\:s^{-1}}$)    & \citet{Roberts23} \\
$M_{pl}$            & 32.59$^{+7.31}_{-7.38}$     & Mass of the planet (M$_\oplus$)           & \citet{Roberts23} \\
$R_{pl}/R_*$            & 0.197$\pm$0.002     & Scaled radius of the planet (R$_*$)           & \citet{Roberts23} \\
$b$                 & 0.089$^{+0.082}_{-0.061}$    & Impact Parameter                          & \citet{Roberts23} \\
a & 0.0361 $\pm$0.0008 & Semimajor axis (au) & \citet{Roberts23} \\
\hline
\end{tabular}
\end{table*}

\begin{figure*}[t!]
\begin{center}
\includegraphics[width=0.9\textwidth]{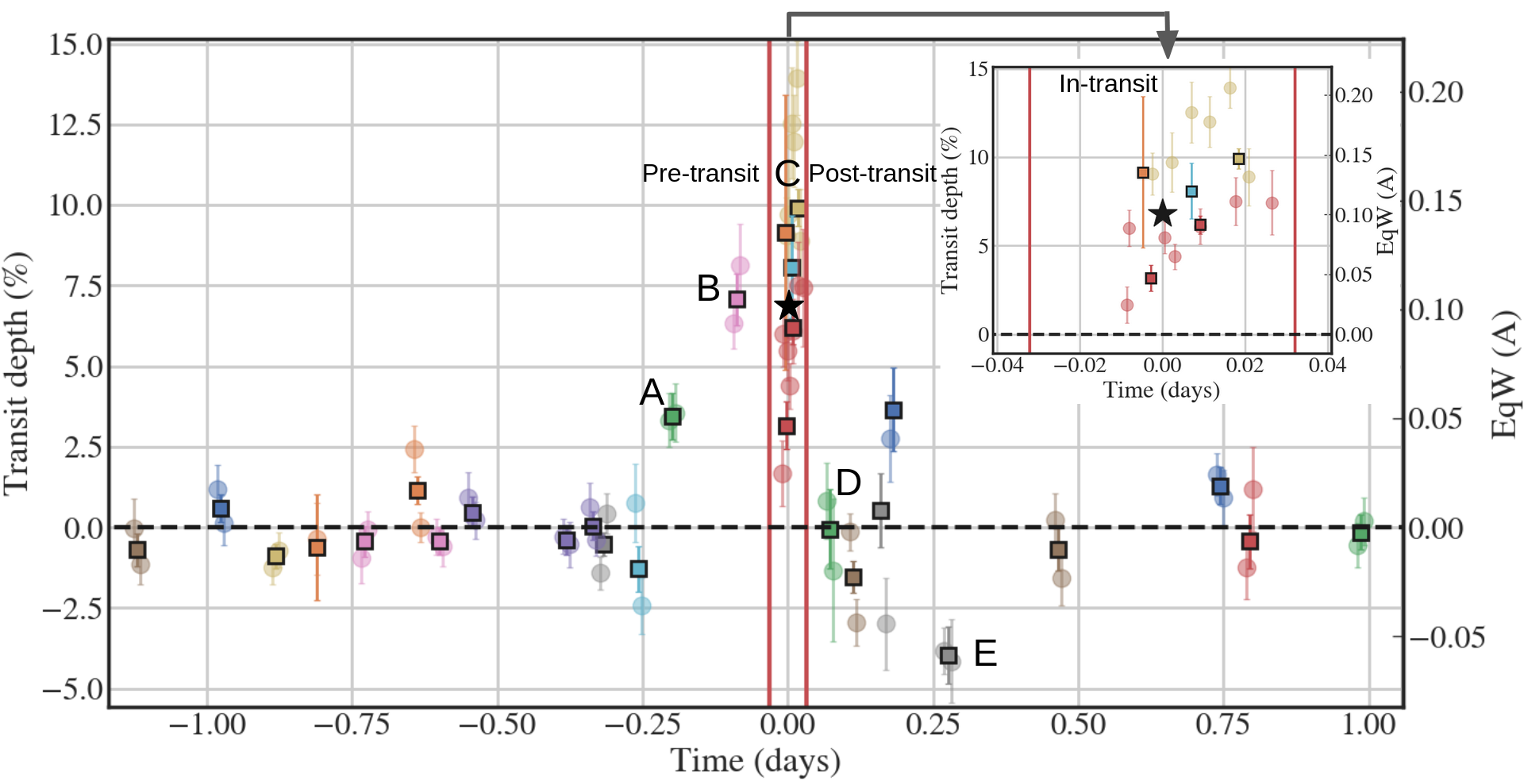}
\caption{ This figure shows the evolution of the relative transit depth and equivalent width of the He I 10830\AA\, absorption in TOI-3884b across the pre-transit (A \& B), in-transit (C), and post-transit phases (D). The outlier E is explained in Section \ref{sec:activity}. The x-axis shows the orbital phase, centered on mid-transit, in units of days. The left y-axis shows the transit depth in percent and the right y-axis shows the scaled equivalent width in \AA . The circular points represent the transit depth measured for individual spectrum with S/N $>$ 20. The square points represent the measured transit depth from the spectra averaged within each visit. Each color is an independent night of observation. The vertical red lines mark the duration of the planet's transit. The inset shows a zoomed-in view of the phase range inside the transit window. The average in-transit signal is marked by a black star symbol in the inset. Since the spectra at each epoch were continuum normalised, the optical transit depth of planet is not shown here.} 
\label{fig:TimeVariableTranistDepth}
\end{center}
\end{figure*}

The transit depth and corresponding scaled equivalent width for all epochs are shown in Figure \ref{fig:TimeVariableTranistDepth}.  

\subsection{Magnetic field measurement}
\label{sec:magfield}
\begin{figure*}[t!]
\begin{center}
\includegraphics[width=2.0\columnwidth]{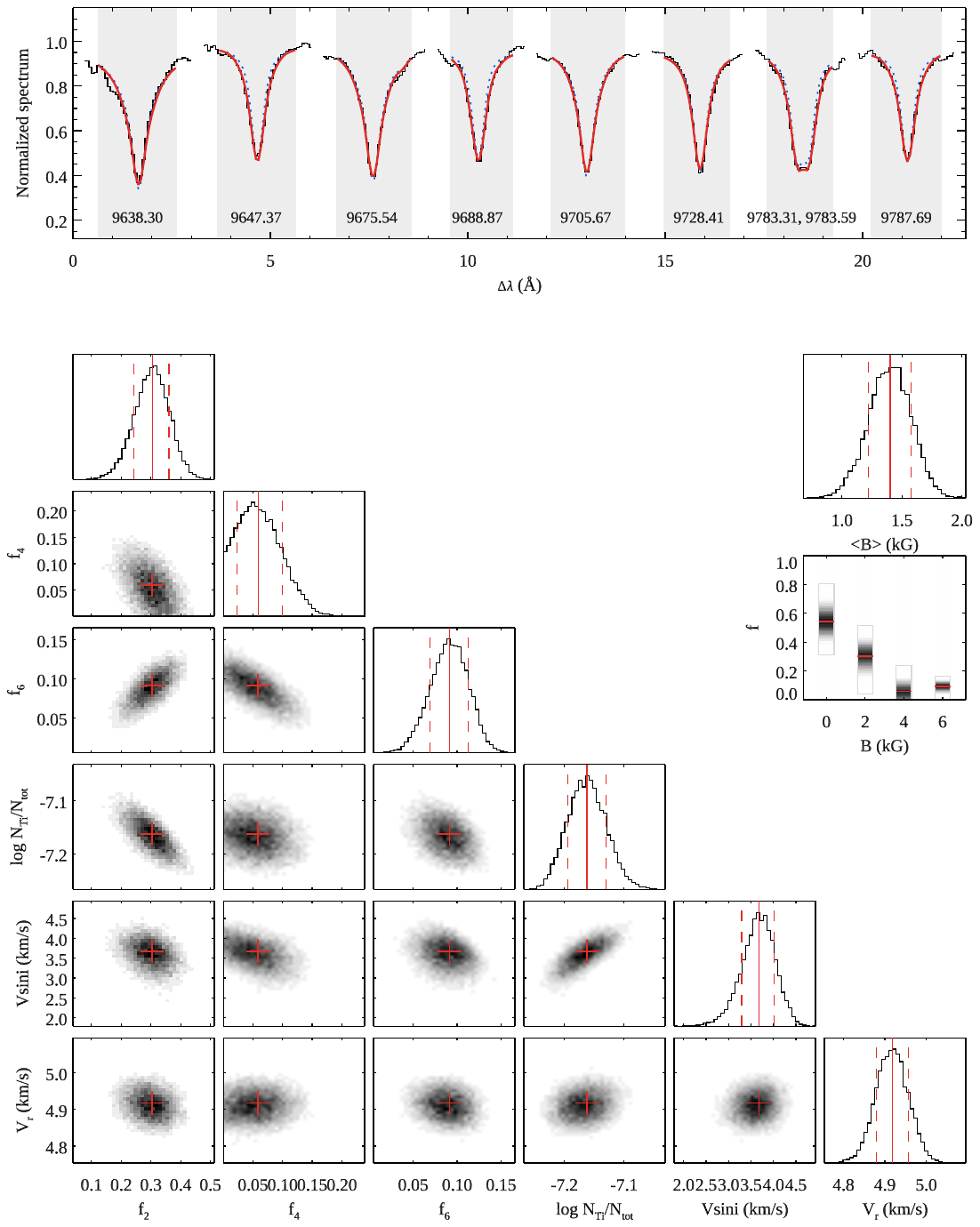}
\caption{This figure shows the best-fit model of Ti I lines in the average HPF spectrum of TOI-3884 and the corresponding corner plot of the model parameters. The absolute velocity V$_r$ term in the plot is at an offset of 3.16 km s$^{-1}$. The marginalized posterior distributions of the filling factors for 0 kG, 2 kG, 4 kG, and 6 kG are also shown on the right. The small increase in the 6 kG filling factor could be due to the active spots on TOI3884.}
\label{fig:magfieldTOI3884}
\end{center}
\end{figure*}

To model the photospheric magnetic field, we modeled a set of eight Ti I lines in the multi-epoch-averaged HPF spectra of TOI-3884 (with a combined signal to noise ratio of $\sim$300). The details of the modeling are described in \citet{Getman25, Amateis26}. Briefly, the spectral synthesis of these magnetically sensitive lines was performed by including the effects of Zeeman broadening and polarized radiative transfer. The observed spectra were modeled as a linear combination of synthetic models with field strengths of 0 kG, 2 kG, 4kG, and 6 kG, with variable filling factors. Figure \ref{fig:magfieldTOI3884} shows the inferred filling factors for different magnetic field strengths. The average field strength obtained for TOI-3884 is $\langle B \rangle =1.40^{+0.17}_ {-0.18}$ kG. The small increase in the filling factor for 6 kG relative to 4 kG (see Figure \ref{fig:magfieldTOI3884}) might be due to the large active spot region.

\section{Results and Discussion} \label{sec:discussion}
\subsection{In-transit Signal}
\label{sec:intranitsignal}
We detect a strong He I 10830\AA\, signature, with an excess transit depth of 6.8$\pm$0.27\% and an equivalent width of 0.101$\pm$0.004\AA\, on weighted average across the five transits we observed. The width of the profile (see Figure \ref{fig:AvgHe10830Profile}), combined with the relatively small change in the planet's velocity during transit, makes it difficult to estimate any velocity variation during the transit.
As detailed in Section \ref{sec:activity}, our confidence in the in-transit signal from the planet is strengthened by the comparable levels of pre-transit signal seen in this system.  
We detect significant variations in the strength of the He I 10830\AA\, signal across the five transits. While the signal was present in all the transits, its amplitude varied from transit to transit (see the scatter in the inset of Figure\ref{fig:TimeVariableTranistDepth}). Further follow-up observations are needed to determine whether this variability is due to the transit light source effect (TLSE) or to fluctuations in the exoplanet's exospheric outflow.

\subsection{1D model}
\begin{figure*}[t!]
\begin{center}
\includegraphics[width=2.0\columnwidth]{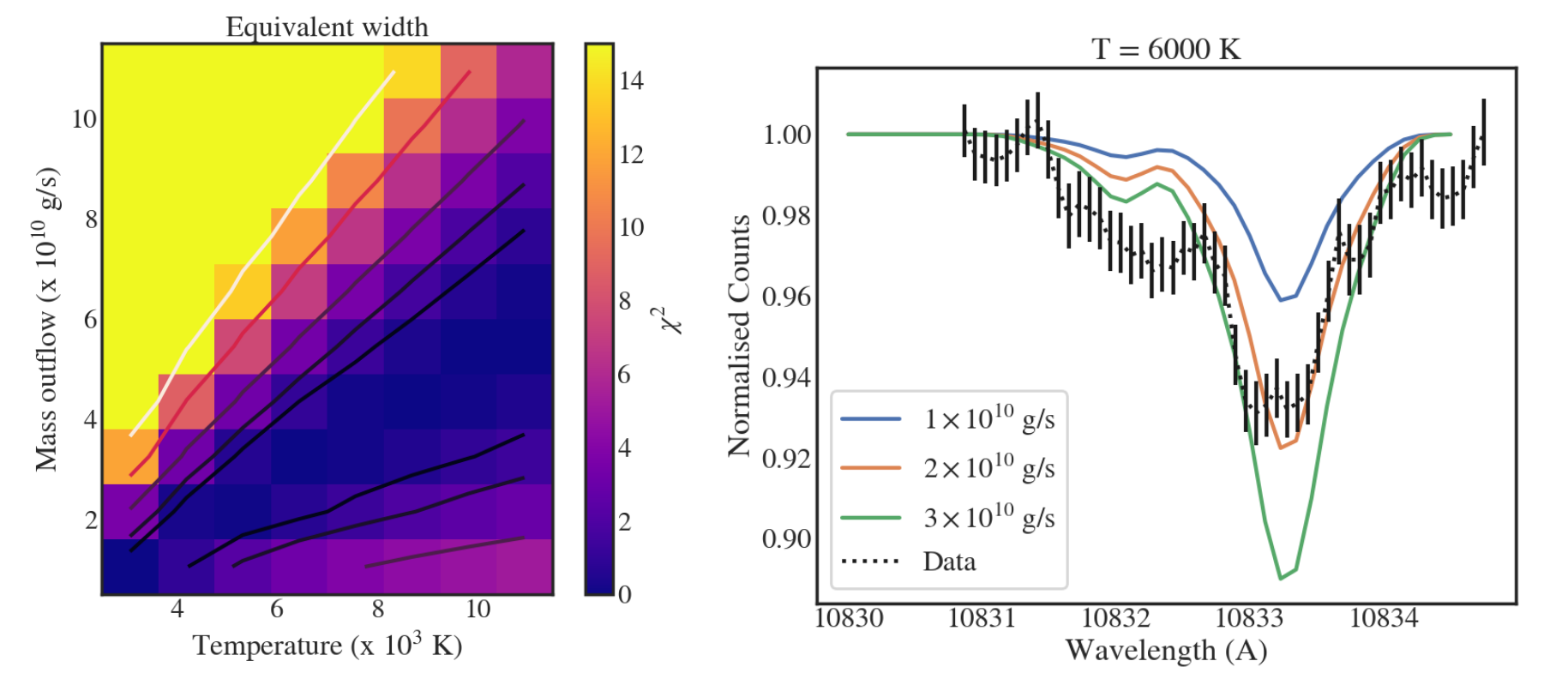}
\caption{Panels (a) and (b) illustrate the 1D Parker-wind model grid. (a) The color map shows the chi-square valley in the  $\dot{M}$-$T$ space for the measured equivalent widths compared to the 1D model equivalent width. The contours of the valley are also overlaid on the image. (b) The colored curves show line profiles predicted for a 1D model at 6000 K for different mass-loss rates, compared to the 3-pixel smoothed observed average line profile (black dotted curve with error bars).}
\label{fig:1dmodelgrid}
\end{center}
\end{figure*}

To constrain the mass outflow rate, we modeled the exospheric outflow using an isothermal 1D Parker wind. While this model cannot explain the pre-transit signature, its simplicity allows us to constrain the mass outflow rate using the equivalent-width measurement during transit. The modeling of metastable helium atoms and the line radiative transfer were carried out following the procedure developed in \citet{Ninan20,oklopcic18}. TOI-3884, the host star, is an M4 dwarf star (P$_{rot}$ $\sim$ 11 days; \citet{Mori25}). The XUV spectrum irradiating the planet is one of the most critical inputs for modeling the exosphere. For instance, our model using the XUV spectrum of another M4 dwarf star, GJ876, from the X-exoplanets archive \citep{Forcada11} produced no He I 10830\AA\, signature. The XUV spectrum of the moderate to low activity GJ 876 (P$_{rot}$ $\sim$ 97 days; \citet{Rivera10})
was insufficient to ionize helium rapidly enough to drive an outflow with any detectable He I 10830\AA\, signature. Since TOI-3884 is an active star with a persistent spot \citep{Roberts23}, we instead adopted the XUV spectrum of an  active M4 dwarf star, YZ CMi  (P$_{rot}$ $\sim$ 2.77 days; \citet{Donati08}), to irradiate the planetary atmosphere in our model (estimated from FUV spectroscopy with HST; GO 17428, PI - K. France).  YZ CMi has a log($R^\prime_{HK}$) = -4.078 \citep{Astudillo17}, while TOI-3884 has a log($R^\prime_{HK}$) = -4.79 $\pm$ 0.13 \citep{Almenara22}. This model produced a significant outflow and a He I 10830\AA\, signature consistent with the observations. 
Since the Parker-wind-based 1D model has two parameters, $\dot{M}$ for the mass-loss rate and $T$ for the temperature of the upper exosphere, we compared models across a grid using the equivalent width measurement. Figure \ref{fig:1dmodelgrid}(a) shows the chi-square valley in the $\dot{M}$-$T$ space that matches the observed equivalent width. The right panel of Figure \ref{fig:1dmodelgrid}(b) shows how the modeled profile compares with the observed data.
Within the caveats of this 1D model, TOI-3884b has an evaporating atmosphere with a mass-loss rate of $\sim 1-4 \times 10^{10}$ g s$^{-1}$ for a typical isothermal model with $T =  4000 - 6000$ K. 

\subsection{Pre-transit}
In two epochs when we observed before the optical transit started, we see a clear pre-transit He I 10830\AA\, absorption signature in the data (see Figure \ref{fig:TimeVariableTranistDepth}). However, we do not see any indication of a post-transit cometary tail signature. As far as we know, the only other system that has shown this behavior is HAT-P-67b \citep{Santiago24}. The host star of HAT-P-67b is an F5 subgiant, and the 3D modeling of HAT-P-67b \citep{Nail25} suggests that this kind of dayside-only outflow can occur if the system is sufficiently cool and the exosphere easily overflows the Roche lobe from the dayside. It is interesting that we see a very similar outflow structure in this planet, which orbits an M4 dwarf star. Another system with a longer pre-transit signal than post-transit tail is HAT-P-32b \citep{Zhang23}. In all three cases, the absorption signal is strong: 10\% in HAT-P-67b, 8.2\% in HAT-P32b, and 6.8\% in TOI-3884b. However, the mass-loss rate in TOI-3884b is three orders of magnitude smaller. Since the inactive photosphere and chromosphere of an M4 dwarf does not produce enough XUV flux to drive an outflow, we hypothesize that the persistent active regions on this host as well as an active chromosphere could be driving this dayside-only outflow. Since a core-powered mass outflow is symmetric, this strong asymmetry detected between the pre- and post-transit signals further supports irradiation driven mass loss over core-powered mass outflow. The pre-transit signature starts $\sim$0.25 days before the optical transit. This implies that the headwind stream extends out to about 23 times the planetary diameter (7.8 times the Roche lobe radius).

\subsection{Bubble versus Stream morphology}
\citet{MacLeod25} proposed a classification of exoplanetary outflow morphology into bubble and stream systems. When the planet has a hot (i.e., faster) outflow, the wind escapes symmetrically and forms a bubble morphology, resulting in a nearly isotropic absorption signature during both pre- and post-transit observations. When the planet has a colder (i.e., slower) outflow, the wind is affected by the two-body gravitational potential beyond the Hill radius, and it primarily escapes as a stream via the L1 and L2 Lagrange points. Stream morphology can lead to the detection of extended pre- or post-transit tails. This morphological distinction can be quantified by the Rossby number. The Roche lobe radius of TOI-3884b is $r_h= 5.8 R_p$ . For our best-fit 1D Parker wind model, the outflow velocity at the Roche lobe is  $v_h = 7.5$ km s$^{-1}$ and the sound speed is $c_s = vs = 7.9$ km s$^{-1}$. This implies a hydrodynamic escape parameter of $\lambda_H = 0.9$ and a Rossby number of $R_{oH} = 1.8$. In the idealized \citet{MacLeod25} model, a Rossby number greater than 1 implies a bubble morphology rather than a stream morphology.
However, similar to the case of HAT-P-67b, despite the slightly higher Rossby number, the pre-transit absorption indicates a stream-like structure. 

\subsection{Magnetically channeled outflow}
\label{sec:magchannel}
When a planet orbits inside a star's Alfvén radius, that is, where the Alfvén speed exceeds the stellar wind speed, the planet will not form a magnetic bow shock; instead, it threads through the stellar magnetic field and forms magnetic-field line channels to the star \citep{Strugarek25arXiv250213262S}. The exosphere outflow can then be channeled along open field lines toward the star through the polar regions \citep{Adams11}. 

Since the dipole dominates at large distances, we assume a $1/r^3$ radial falloff of the field strength starting with $\langle B \rangle =1.40^{+0.17}_ {-0.18}$ kG at the photosphere estimated in Section \ref{sec:magfield}. This implies a field strength of $B = 8 \times 10^{-5}$ kG at the orbital distance of 25.9 R$_*$ for TOI-3884b.
The density of the stellar wind depends on the mass outflow of the host star. Mass outflow rates from M-dwarf stars are poorly constrained. From mass conservation arguments, the density scales as the inverse square of radius $1/r^2$, and inversely with velocity $1/v$. In an isothermal Parker wind or Alfvén wave driven wind, the velocity increase is nominal outside the critical point. We approximate the $1/r^2$ regime in the outer regions of the stellar wind density profile presented in Figure 2 of \citet{Shoda20} for Sun-like stars, with $\sim 1 \times 10^{-18}/r^{2}$ $g~ cm^{-3}$ , where $r$ is the orbital radius in units of stellar radii. This implies a mass density of $\rho \sim 1.49 \times 10^{-21}$ $g~ cm^{-3}$ in the stellar wind at TOI-3884b's orbital distance. The stellar wind velocity at this distance is $\sim 400 ~km~ s^{-1}$ \citep{Shoda20}. The Alfvén speed is given by the formula $\frac{B(r)}{\sqrt{\mu_0 \rho(r)}}$, where $\mu_0$ is the permeability of the vacuum. Hence, at the planet's location, the Alfvén speed is $\sim 5887 ~km~ s^{-1}$. This Alfvén speed is significantly greater than the stellar wind speed at TOI-3884b, and therefore we expect sub-Alfvénic  interaction between the planet and the star.

Since TOI-3884b is in a polar orbit (see Figure 9 in \citet{Roberts23}), the sub-Alfvénic interaction opens up an alternate hypothesis for the pre-transit signature. It could be a magnetically channeled outflow. The planet's magnetic fields could connect to the polar region of the stellar hemisphere visible to us. Before transit begins, this funnel connecting the planet to the polar region could be channeling the outflow whose signature we detect in He I 10830 \AA\,. After the transit, this channel will be weaker as the planet moves away from this pole, and toward the far hemisphere pole behind the star, resulting in the lack of a post-transit tail. 

A complete MHD simulation of this interaction is beyond the scope of this manuscript, but it remains a viable hypothesis to explore.

\subsection{Stellar activity contamination}
\label{sec:activity}
Since TOI-3884b transits consistently over the active spot region \citep{Roberts23}, the observations could be susceptible to the transit light source effect (TLSE), i.e., if there are bright He I 10830\AA\,-emitting regions around these stellar spots that are selectively blocked by the planet during transit, it can result in an extra absorption dip that masquerades as the planetary signal.  We could not rule out the TLSE using velocity variations during transit, because of the small expected velocity shifts and the insufficient S/N. If we assume that the active regions correspond to the three spots estimated in \citet{Roberts23}, then for the planet to produce a 6.8\% extra transit depth, the excess He I emission per unit area in the active regions must be at least 4 times that of the rest of the stellar disc. We can confidently rule out the hypothesis that the detected signal is entirely due to TLSE, since that scenario cannot explain the strong pre-transit absorption signal. However, we cannot rule out some level of stellar contamination in the amplitude of the in-transit signal or its variations reported in Section \ref{sec:intranitsignal}. 
One of the out-of-transit outliers in Figure \ref{fig:TimeVariableTranistDepth}, epoch marked by E, can be explained by a strong stellar activity event. At that epoch spectrum, we could clearly see a change in stellar rest frame He 10830 \AA\, line strength.

\section{Conclusion}\label{sec:conclusion}
We report the detection of a strong exospheric outflow from TOI-3884b, despite its host star being an M4 dwarf. In our isothermal 4000 - 6000 K model, the planet is losing mass at a rate of $\sim 1 - 4  \times 10^{10}$ g s$^{-1}$. Since XUV radiation from an inactive M4 star such as GJ 876 cannot drive the atmospheric outflow or ionize Helium efficiently enough to produce metastable helium atoms, we hypothesize that the detected outflow is driven by XUV radiation from chromospheric active regions on the TOI-3884 host star, analogous to those in YZ Cmi.  The combination of a detected pre-transit headwind and the lack of a post-transit tail is unusual. Its behavior is similar to that of HAT-P-67b, a hot Saturn orbiting an F5 dwarf star. TOI-3884b is orbiting a much cooler M4 dwarf. Such an extended headwind, which is longer than $\sim$23 times the planetary diameter, will need to be explained by advanced exospheric models of star-planet interactions. These discoveries also caution against using spectra taken close to the transit as out-of-transit baselines, since the exosphere can be many times larger than the exoplanet. 


\begin{acknowledgments}
J.P.N. \& S.M. thank Harish Vedantham and Sagnik Saha for discussion on the sub-Alfvénic interaction at ICTS-TIFR Strange New Worlds School in 2026 April.
The Center for Exoplanets and Habitable Worlds is supported by the Pennsylvania State University, the Eberly College of Science, and the Pennsylvania Space Grant Consortium. 
Computations for this research were performed on the Pennsylvania State University’s Institute for CyberScience Advanced CyberInfrastructure (ICS-ACI).

These results are based on observations obtained with the Habitable-zone Planet Finder Spectrograph on the Hobby-Eberly Telescope. We thank the Resident astronomers and Telescope Operators at the HET for the skillful execution of our observations of our observations with HPF. The Hobby-Eberly Telescope is a joint project of the University of Texas at Austin, the Pennsylvania State University, Ludwig-Maximilians-Universität München, and Georg-August Universität Gottingen. The HET is named in honor of its principal benefactors, William P. Hobby and Robert E. Eberly. The HET collaboration acknowledges the support and resources from the Texas Advanced Computing Center. 

S.P.H acknowledges that this research was carried out, in part, at the Jet Propulsion Laboratory, California Institute of Technology, under a contract with the National Aeronautics and Space Administration (80NM0018D0004). 

This research has made use of NASA's Astrophysics Data System Bibliographic Services. Perplexity AI was used for checking grammar and getting critical feedback on language quality. After using this tool/service, the author(s) reviewed and edited the content as needed and take(s) full responsibility for the content of the publication.

J.P.N. acknowledge the support of the Department of Atomic Energy, Government of India, under Project Identification No. RTI 4002. J.P.N acknowledges support in part by a generous donation (from the Murty Trust) aimed at enabling advances in astrophysics through the use of machine learning. The Murty Trust, an initiative of the Murty Foundation, is a not-for-profit organization dedicated to the preservation and celebration of culture, science, and knowledge of systems born out of India. The Murty Trust is headed by Mrs. Sudha Murty and Mr. Rohan Murty.
O.K. acknowledges support by the Swedish Research Council (grant agreement no. 2023-03667) and by the Swedish National Space Agency. S.M, M.D., P.R. acknowledge support from NASA through the Exoplanet (XRP) research program grant 80NSSC24K0155. S.M., P.R., G.S. and J.L-R. are part of NASA's CHAMPs team, supported by NASA under grant No. 80NSSC23K1399 issued through the Interdisciplinary Consortia for Astrobiology Research (ICAR) program.

\end{acknowledgments}

%

\vspace{5mm}
\facilities{HPF(HET)}


\software{HxRGProc \citep{ninan18}, 
  barycorrpy \citep{kanodia18},
  astropy \citep{astropy:2013, astropy:2018, astropy:2022}, numpy \citep{2020NumPy-Array}, scipy \citep{2020SciPy-NMeth}, matplotlib \citep{Hunter2007}, GNU parallel \citep{Tange2011a}
}



\bibliography{HPFreferences}

@INPROCEEDINGS{HETRamsey,
   author = {{Ramsey}, L.~W. and {Adams}, M.~T. and {Barnes}, T.~G. and {Booth}, J.~A. and 
	{Cornell}, M.~E. and {Fowler}, J.~R. and {Gaffney}, N.~I. and 
	{Glaspey}, J.~W. and {Good}, J.~M. and {Hill}, G.~J. and {Kelton}, P.~W. and 
	{Krabbendam}, V.~L. and {Long}, L. and {MacQueen}, P.~J. and 
	{Ray}, F.~B. and {Ricklefs}, R.~L. and {Sage}, J. and {Sebring}, T.~A. and 
	{Spiesman}, W.~J. and {Steiner}, M.},
    title = "{Early performance and present status of the Hobby-Eberly Telescope}",
booktitle = {Advanced Technology Optical/IR Telescopes VI},
     year = 1998,
   series = {\procspie},
   volume = 3352,
   editor = {{Stepp}, L.~M.},
    month = aug,
    pages = {34-42},
      doi = {10.1117/12.319287},
   adsurl = {http://adsabs.harvard.edu/abs/1998SPIE.3352...34R}
}

@ARTICLE{HETqueue,
   author = {{Shetrone}, M. and {Cornell}, M.~E. and {Fowler}, J.~R. and 
	{Gaffney}, N. and {Laws}, B. and {Mader}, J. and {Mason}, C. and 
	{Odewahn}, S. and {Roman}, B. and {Rostopchin}, S. and {Schneider}, D.~P. and 
	{Umbarger}, J. and {Westfall}, A.},
    title = "{Ten Year Review of Queue Scheduling of the Hobby-Eberly Telescope}",
  journal = {\pasp},
archivePrefix = "arXiv",
   eprint = {0705.3889},
     year = 2007,
    month = may,
   volume = 119,
    pages = {556-566},
      doi = {10.1086/519291},
   adsurl = {http://adsabs.harvard.edu/abs/2007PASP..119..556S}
}

@INPROCEEDINGS{mah2014SPIE,
   author = {{Mahadevan}, S. and {Ramsey}, L.~W. and {Terrien}, R. and {Halverson}, S. and 
	{Roy}, A. and {Hearty}, F. and {Levi}, E. and {Stefansson}, G.~K. and 
	{Robertson}, P. and {Bender}, C. and {Schwab}, C. and {Nelson}, M.
	},
    title = "{The Habitable-zone Planet Finder: A status update on the development of a stabilized fiber-fed near-infrared spectrograph for the for the Hobby-Eberly telescope}",
booktitle = {Ground-based and Airborne Instrumentation for Astronomy V},
     year = 2014,
   series = {\procspie},
   volume = 9147,
    month = jul,
      eid = {91471G},
    pages = {91471G},
      doi = {10.1117/12.2056417},
   adsurl = {http://adsabs.harvard.edu/abs/2014SPIE.9147E..1GM}
}

@INPROCEEDINGS{ninan18,
   author = {{Ninan}, J.~P. and {Bender}, C.~F. and {Mahadevan}, S. and {Ford}, E.~B. and 
	{Monson}, A.~J. and {Kaplan}, K.~F. and {Terrien}, R.~C. and 
	{Roy}, A. and {Robertson}, P.~M. and {Kanodia}, S. and {Stefansson}, G.~K.
	},
    title = "{The Habitable-Zone Planet Finder: improved flux image generation algorithms for H2RG up-the-ramp data}",
booktitle = {High Energy, Optical, and Infrared Detectors for Astronomy VIII},
     year = 2018,
   series = {Society of Photo-Optical Instrumentation Engineers (SPIE) Conference Series},
   journal = {SPIE},
   volume = 10709,
    month = aug,
      eid = {107092U},
    pages = {107092U},
      doi = {10.1117/12.2312787},
   adsurl = {https://ui.adsabs.harvard.edu/abs/2018SPIE10709E..2UN}
}

@article{ehrenreich15,
	title = {A giant comet-like cloud of hydrogen escaping the warm {Neptune}-mass exoplanet {GJ} 436b},
	volume = {522},
	copyright = {2015 Nature Publishing Group},
	issn = {1476-4687},
	url = {https://www.nature.com/articles/nature14501},
	doi = {10.1038/nature14501},
	language = {en},
	number = {7557},
	urldate = {2018-03-14},
	journal = {Nature},
	author = {Ehrenreich, David and Bourrier, Vincent and Wheatley, Peter J. and des Etangs, Alain Lecavelier and Hébrard, Guillaume and Udry, Stéphane and Bonfils, Xavier and Delfosse, Xavier and Désert, Jean-Michel and Sing, David K. and Vidal-Madjar, Alfred},
	month = jun,
	year = {2015},
	pages = {459--461}
}

@article{vidalmadjar03,
	title = {An extended upper atmosphere around the extrasolar planet {HD}209458b},
	volume = {422},
	copyright = {2003 Nature Publishing Group},
	issn = {1476-4687},
	url = {https://www.nature.com/articles/nature01448},
	doi = {10.1038/nature01448},
	language = {en},
	number = {6928},
	urldate = {2018-03-14},
	journal = {Nature},
	author = {Vidal-Madjar, A. and des Etangs, A. Lecavelier and Désert, J.-M. and Ballester, G. E. and Ferlet, R. and Hébrard, G. and Mayor, M.},
	month = mar,
	year = {2003},
	pages = {143--146}
}

@article{oklopcic18,
	title = {A {New} {Window} into {Escaping} {Exoplanet} {Atmospheres}: 10830 Å {Line} of {Helium}},
	volume = {855},
	issn = {2041-8205},
	shorttitle = {A {New} {Window} into {Escaping} {Exoplanet} {Atmospheres}},
	url = {http://stacks.iop.org/2041-8205/855/i=1/a=L11},
	doi = {10.3847/2041-8213/aaada9},
	language = {en},
	number = {1},
	urldate = {2018-03-14},
	journal = {The Astrophysical Journal Letters},
	author = {Oklop{\v{c}}i{\'c}, Antonija and Hirata, Christopher M.},
	year = {2018},
	pages = {L11}
}

@article{seager00,
	title = {Theoretical {Transmission} {Spectra} during {Extrasolar} {Giant} {Planet} {Transits}},
	volume = {537},
	issn = {0004-637X},
	url = {http://stacks.iop.org/0004-637X/537/i=2/a=916},
	doi = {10.1086/309088},
	language = {en},
	number = {2},
	urldate = {2018-03-15},
	journal = {The Astrophysical Journal},
	author = {Seager, S. and Sasselov, D. D.},
	year = {2000},
	pages = {916}
}

@ARTICLE{kulow14,
   author = {{Kulow}, J.~R. and {France}, K. and {Linsky}, J. and {Loyd}, R.~O.~P.
	},
    title = "{Ly{$\alpha$} Transit Spectroscopy and the Neutral Hydrogen Tail of the Hot Neptune GJ 436b}",
  journal = {\apj},
archivePrefix = "arXiv",
   eprint = {1403.6834},
 primaryClass = "astro-ph.EP",
     year = 2014,
    month = may,
   volume = 786,
      eid = {132},
    pages = {132},
      doi = {10.1088/0004-637X/786/2/132},
   adsurl = {https://ui.adsabs.harvard.edu/abs/2014ApJ...786..132K}
}

@ARTICLE{bourrier18,
       author = {{Bourrier}, V. and {Lecavelier des Etangs}, A. and {Ehrenreich}, D. and
         {Sanz-Forcada}, J. and {Allart}, R. and {Ballester}, G.~E. and
         {Buchhave}, L.~A. and {Cohen}, O. and {Deming}, D. and {Evans}, T.~M. and
         {Garc{\'\i}a Mu{\~n}oz}, A. and {Henry}, G.~W. and {Kataria}, T. and
         {Lavvas}, P. and {Lewis}, N. and {L{\'o}pez-Morales}, M. and
         {Marley}, M. and {Sing}, D.~K. and {Wakeford}, H.~R.},
        title = "{Hubble PanCET: an extended upper atmosphere of neutral hydrogen around the warm Neptune GJ 3470b}",
      journal = {\aap},
         year = "2018",
        month = "Dec",
       volume = {620},
          eid = {A147},
        pages = {A147},
          doi = {10.1051/0004-6361/201833675},
archivePrefix = {arXiv},
       eprint = {1812.05119},
 primaryClass = {astro-ph.EP},
       adsurl = {https://ui.adsabs.harvard.edu/abs/2018A&A...620A.147B}
}

@ARTICLE{spake18,
   author = {{Spake}, J.~J. and {Sing}, D.~K. and {Evans}, T.~M. and {Oklop{\v c}i{\'c}}, A. and 
	{Bourrier}, V. and {Kreidberg}, L. and {Rackham}, B.~V. and 
	{Irwin}, J. and {Ehrenreich}, D. and {Wyttenbach}, A. and {Wakeford}, H.~R. and 
	{Zhou}, Y. and {Chubb}, K.~L. and {Nikolov}, N. and {Goyal}, J.~M. and 
	{Henry}, G.~W. and {Williamson}, M.~H. and {Blumenthal}, S. and 
	{Anderson}, D.~R. and {Hellier}, C. and {Charbonneau}, D. and 
	{Udry}, S. and {Madhusudhan}, N.},
    title = "{Helium in the eroding atmosphere of an exoplanet}",
  journal = {\nat},
archivePrefix = "arXiv",
   eprint = {1805.01298},
 primaryClass = "astro-ph.EP",
     year = 2018,
    month = may,
   volume = 557,
    pages = {68-70},
      doi = {10.1038/s41586-018-0067-5},
   adsurl = {https://ui.adsabs.harvard.edu/abs/2018Natur.557...68S}
}

@ARTICLE{allart18,
   author = {{Allart}, R. and {Bourrier}, V. and {Lovis}, C. and {Ehrenreich}, D. and 
	{Spake}, J.~J. and {Wyttenbach}, A. and {Pino}, L. and {Pepe}, F. and 
	{Sing}, D.~K. and {Lecavelier des Etangs}, A.},
    title = "{Spectrally resolved helium absorption from the extended atmosphere of a warm Neptune-mass exoplanet}",
  journal = {Science},
archivePrefix = "arXiv",
   eprint = {1812.02189},
 primaryClass = "astro-ph.EP",
     year = 2018,
    month = dec,
   volume = 362,
    pages = {1384-1387},
      doi = {10.1126/science.aat5879},
   adsurl = {https://ui.adsabs.harvard.edu/abs/2018Sci...362.1384A}
}

@ARTICLE{allart19,
   author = {{Allart}, R. and {Bourrier}, V. and {Lovis}, C. and {Ehrenreich}, D. and 
	{Aceituno}, J. and {Guijarro}, A. and {Pepe}, F. and {Sing}, D.~K. and 
	{Spake}, J.~J. and {Wyttenbach}, A.},
    title = "{High-resolution confirmation of an extended helium atmosphere around WASP-107b}",
  journal = {\aap},
archivePrefix = "arXiv",
   eprint = {1901.08073},
 primaryClass = "astro-ph.EP",
     year = 2019,
    month = mar,
   volume = 623,
      eid = {A58},
    pages = {A58},
      doi = {10.1051/0004-6361/201834917},
   adsurl = {https://ui.adsabs.harvard.edu/abs/2019A%26A...623A..58A}
}

@ARTICLE{nortmann18,
   author = {{Nortmann}, L. and {Pall{\'e}}, E. and {Salz}, M. and {Sanz-Forcada}, J. and 
	{Nagel}, E. and {Alonso-Floriano}, F.~J. and {Czesla}, S. and 
	{Yan}, F. and {Chen}, G. and {Snellen}, I.~A.~G. and {Zechmeister}, M. and 
	{Schmitt}, J.~H.~M.~M. and {L{\'o}pez-Puertas}, M. and {Casasayas-Barris}, N. and 
	{Bauer}, F.~F. and {Amado}, P.~J. and {Caballero}, J.~A. and 
	{Dreizler}, S. and {Henning}, T. and {Lamp{\'o}n}, M. and {Montes}, D. and 
	{Molaverdikhani}, K. and {Quirrenbach}, A. and {Reiners}, A. and 
	{Ribas}, I. and {S{\'a}nchez-L{\'o}pez}, A. and {Schneider}, P.~C. and 
	{Zapatero Osorio}, M.~R.},
    title = "{Ground-based detection of an extended helium atmosphere in the Saturn-mass exoplanet WASP-69b}",
  journal = {Science},
archivePrefix = "arXiv",
   eprint = {1812.03119},
 primaryClass = "astro-ph.EP",
     year = 2018,
    month = dec,
   volume = 362,
    pages = {1388-1391},
      doi = {10.1126/science.aat5348},
   adsurl = {https://ui.adsabs.harvard.edu/abs/2018Sci...362.1388N}
}

@ARTICLE{mansfield18,
   author = {{Mansfield}, M. and {Bean}, J.~L. and {Oklop{\v c}i{\'c}}, A. and 
	{Kreidberg}, L. and {D{\'e}sert}, J.-M. and {Kempton}, E.~M.-R. and 
	{Line}, M.~R. and {Fortney}, J.~J. and {Henry}, G.~W. and {Mallonn}, M. and 
	{Stevenson}, K.~B. and {Dragomir}, D. and {Allart}, R. and {Bourrier}, V.
	},
    title = "{Detection of Helium in the Atmosphere of the Exo-Neptune HAT-P-11b}",
  journal = {\apjl},
archivePrefix = "arXiv",
   eprint = {1812.02214},
 primaryClass = "astro-ph.EP",
     year = 2018,
    month = dec,
   volume = 868,
      eid = {L34},
    pages = {L34},
      doi = {10.3847/2041-8213/aaf166},
   adsurl = {https://ui.adsabs.harvard.edu/abs/2018ApJ...868L..34M}
}

@ARTICLE{salz18,
   author = {{Salz}, M. and {Czesla}, S. and {Schneider}, P.~C. and {Nagel}, E. and 
	{Schmitt}, J.~H.~M.~M. and {Nortmann}, L. and {Alonso-Floriano}, F.~J. and 
	{L{\'o}pez-Puertas}, M. and {Lamp{\'o}n}, M. and {Bauer}, F.~F. and 
	{Snellen}, I.~A.~G. and {Pall{\'e}}, E. and {Caballero}, J.~A. and 
	{Yan}, F. and {Chen}, G. and {Sanz-Forcada}, J. and {Amado}, P.~J. and 
	{Quirrenbach}, A. and {Ribas}, I. and {Reiners}, A. and {B{\'e}jar}, V.~J.~S. and 
	{Casasayas-Barris}, N. and {Cort{\'e}s-Contreras}, M. and {Dreizler}, S. and 
	{Guenther}, E.~W. and {Henning}, T. and {Jeffers}, S.~V. and 
	{Kaminski}, A. and {K{\"u}rster}, M. and {Lafarga}, M. and {Lara}, L.~M. and 
	{Molaverdikhani}, K. and {Montes}, D. and {Morales}, J.~C. and 
	{S{\'a}nchez-L{\'o}pez}, A. and {Seifert}, W. and {Zapatero Osorio}, M.~R. and 
	{Zechmeister}, M.},
    title = "{Detection of He I {$\lambda$}10830 {\AA} absorption on HD 189733 b with CARMENES high-resolution transmission spectroscopy}",
  journal = {\aap},
archivePrefix = "arXiv",
   eprint = {1812.02453},
 primaryClass = "astro-ph.EP",
     year = 2018,
    month = dec,
   volume = 620,
      eid = {A97},
    pages = {A97},
      doi = {10.1051/0004-6361/201833694},
   adsurl = {https://ui.adsabs.harvard.edu/abs/2018A%26A...620A..97S}
}

@ARTICLE{2020SciPy-NMeth,
  author  = {Virtanen, Pauli and Gommers, Ralf and Oliphant, Travis E. and
            Haberland, Matt and Reddy, Tyler and Cournapeau, David and
            Burovski, Evgeni and Peterson, Pearu and Weckesser, Warren and
            Bright, Jonathan and {van der Walt}, St{\'e}fan J. and
            Brett, Matthew and Wilson, Joshua and Millman, K. Jarrod and
            Mayorov, Nikolay and Nelson, Andrew R. J. and Jones, Eric and
            Kern, Robert and Larson, Eric and Carey, C J and
            Polat, {\.I}lhan and Feng, Yu and Moore, Eric W. and
            {VanderPlas}, Jake and Laxalde, Denis and Perktold, Josef and
            Cimrman, Robert and Henriksen, Ian and Quintero, E. A. and
            Harris, Charles R. and Archibald, Anne M. and
            Ribeiro, Ant{\^o}nio H. and Pedregosa, Fabian and
            {van Mulbregt}, Paul and {SciPy 1.0 Contributors}},
  title   = {{{SciPy} 1.0: Fundamental Algorithms for Scientific
            Computing in Python}},
  journal = {Nature Methods},
  year    = {2020},
  volume  = {17},
  pages   = {261--272},
  adsurl  = {https://rdcu.be/b08Wh},
  doi     = {10.1038/s41592-019-0686-2},
}

@ARTICLE{2020NumPy-Array,
  author  = {Harris, Charles R. and Millman, K. Jarrod and
            van der Walt, Stéfan J and Gommers, Ralf and
            Virtanen, Pauli and Cournapeau, David and
            Wieser, Eric and Taylor, Julian and Berg, Sebastian and
            Smith, Nathaniel J. and Kern, Robert and Picus, Matti and
            Hoyer, Stephan and van Kerkwijk, Marten H. and
            Brett, Matthew and Haldane, Allan and
            Fernández del Río, Jaime and Wiebe, Mark and
            Peterson, Pearu and Gérard-Marchant, Pierre and
            Sheppard, Kevin and Reddy, Tyler and Weckesser, Warren and
            Abbasi, Hameer and Gohlke, Christoph and
            Oliphant, Travis E.},
  title   = {Array programming with {NumPy}},
  journal = {Nature},
  year    = {2020},
  volume  = {585},
  pages   = {357–362},
  doi     = {10.1038/s41586-020-2649-2}
}

@article{astropy:2013,
        Adsurl = {https://adsabs.harvard.edu/abs/2013A%26A...558A..33A},
        Archiveprefix = {arXiv},
        Author = {{Astropy Collaboration} and {Robitaille}, T.~P. and {Tollerud}, E.~J. and {Greenfield}, P. and {Droettboom}, M. and {Bray}, E. and {Aldcroft}, T. and {Davis}, M. and {Ginsburg}, A. and {Price-Whelan}, A.~M. and {Kerzendorf}, W.~E. and {Conley}, A. and {Crighton}, N. and {Barbary}, K. and {Muna}, D. and {Ferguson}, H. and {Grollier}, F. and {Parikh}, M.~M. and {Nair}, P.~H. and {Unther}, H.~M. and {Deil}, C. and {Woillez}, J. and {Conseil}, S. and {Kramer}, R. and {Turner}, J.~E.~H. and {Singer}, L. and {Fox}, R. and {Weaver}, B.~A. and {Zabalza}, V. and {Edwards}, Z.~I. and {Azalee Bostroem}, K. and {Burke}, D.~J. and {Casey}, A.~R. and {Crawford}, S.~M. and {Dencheva}, N. and {Ely}, J. and {Jenness}, T. and {Labrie}, K. and {Lim}, P.~L. and {Pierfederici}, F. and {Pontzen}, A. and {Ptak}, A. and {Refsdal}, B. and {Servillat}, M. and {Streicher}, O.},
        Doi = {10.1051/0004-6361/201322068},
        Eid = {A33},
        Eprint = {1307.6212},
        Journal = {\aap},
        Month = oct,
        Pages = {A33},
        Primaryclass = {astro-ph.IM},
        Title = {{Astropy: A community Python package for astronomy}},
        Volume = 558,
        Year = 2013}

@ARTICLE{astropy:2018,
               author = {{Astropy Collaboration} and {Price-Whelan}, A.~M. and
                 {Sip{\H{o}}cz}, B.~M. and {G{\"u}nther}, H.~M. and {Lim}, P.~L. and
                 {Crawford}, S.~M. and {Conseil}, S. and {Shupe}, D.~L. and
                 {Craig}, M.~W. and {Dencheva}, N. and {Ginsburg}, A. and {Vand
                erPlas}, J.~T. and {Bradley}, L.~D. and {P{\'e}rez-Su{\'a}rez}, D. and
                 {de Val-Borro}, M. and {Aldcroft}, T.~L. and {Cruz}, K.~L. and
                 {Robitaille}, T.~P. and {Tollerud}, E.~J. and {Ardelean}, C. and
                 {Babej}, T. and {Bach}, Y.~P. and {Bachetti}, M. and {Bakanov}, A.~V. and
                 {Bamford}, S.~P. and {Barentsen}, G. and {Barmby}, P. and
                 {Baumbach}, A. and {Berry}, K.~L. and {Biscani}, F. and {Boquien}, M. and
                 {Bostroem}, K.~A. and {Bouma}, L.~G. and {Brammer}, G.~B. and
                 {Bray}, E.~M. and {Breytenbach}, H. and {Buddelmeijer}, H. and
                 {Burke}, D.~J. and {Calderone}, G. and {Cano Rodr{\'\i}guez}, J.~L. and
                 {Cara}, M. and {Cardoso}, J.~V.~M. and {Cheedella}, S. and {Copin}, Y. and
                 {Corrales}, L. and {Crichton}, D. and {D'Avella}, D. and {Deil}, C. and
                 {Depagne}, {\'E}. and {Dietrich}, J.~P. and {Donath}, A. and
                 {Droettboom}, M. and {Earl}, N. and {Erben}, T. and {Fabbro}, S. and
                 {Ferreira}, L.~A. and {Finethy}, T. and {Fox}, R.~T. and
                 {Garrison}, L.~H. and {Gibbons}, S.~L.~J. and {Goldstein}, D.~A. and
                 {Gommers}, R. and {Greco}, J.~P. and {Greenfield}, P. and
                 {Groener}, A.~M. and {Grollier}, F. and {Hagen}, A. and {Hirst}, P. and
                 {Homeier}, D. and {Horton}, A.~J. and {Hosseinzadeh}, G. and {Hu}, L. and
                 {Hunkeler}, J.~S. and {Ivezi{\'c}}, {\v{Z}}. and {Jain}, A. and
                 {Jenness}, T. and {Kanarek}, G. and {Kendrew}, S. and {Kern}, N.~S. and
                 {Kerzendorf}, W.~E. and {Khvalko}, A. and {King}, J. and {Kirkby}, D. and
                 {Kulkarni}, A.~M. and {Kumar}, A. and {Lee}, A. and {Lenz}, D. and
                 {Littlefair}, S.~P. and {Ma}, Z. and {Macleod}, D.~M. and
                 {Mastropietro}, M. and {McCully}, C. and {Montagnac}, S. and
                 {Morris}, B.~M. and {Mueller}, M. and {Mumford}, S.~J. and {Muna}, D. and
                 {Murphy}, N.~A. and {Nelson}, S. and {Nguyen}, G.~H. and
                 {Ninan}, J.~P. and {N{\"o}the}, M. and {Ogaz}, S. and {Oh}, S. and
                 {Parejko}, J.~K. and {Parley}, N. and {Pascual}, S. and {Patil}, R. and
                 {Patil}, A.~A. and {Plunkett}, A.~L. and {Prochaska}, J.~X. and
                 {Rastogi}, T. and {Reddy Janga}, V. and {Sabater}, J. and
                 {Sakurikar}, P. and {Seifert}, M. and {Sherbert}, L.~E. and
                 {Sherwood-Taylor}, H. and {Shih}, A.~Y. and {Sick}, J. and
                 {Silbiger}, M.~T. and {Singanamalla}, S. and {Singer}, L.~P. and
                 {Sladen}, P.~H. and {Sooley}, K.~A. and {Sornarajah}, S. and
                 {Streicher}, O. and {Teuben}, P. and {Thomas}, S.~W. and
                 {Tremblay}, G.~R. and {Turner}, J.~E.~H. and {Terr{\'o}n}, V. and
                 {van Kerkwijk}, M.~H. and {de la Vega}, A. and {Watkins}, L.~L. and
                 {Weaver}, B.~A. and {Whitmore}, J.~B. and {Woillez}, J. and
                 {Zabalza}, V. and {Astropy Contributors}},
                title = "{The Astropy Project: Building an Open-science Project and Status of the v2.0 Core Package}",
              journal = {\aj},
                 year = 2018,
                month = sep,
               volume = {156},
               number = {3},
                  eid = {123},
                pages = {123},
                  doi = {10.3847/1538-3881/aabc4f},
        archivePrefix = {arXiv},
               eprint = {1801.02634},
         primaryClass = {astro-ph.IM},
               adsurl = {https://ui.adsabs.harvard.edu/abs/2018AJ....156..123A}
        }

@ARTICLE{astropy:2022,
               author = {{Astropy Collaboration} and {Price-Whelan}, Adrian M. and {Lim}, Pey Lian and {Earl}, Nicholas and {Starkman}, Nathaniel and {Bradley}, Larry and {Shupe}, David L. and {Patil}, Aarya A. and {Corrales}, Lia and {Brasseur}, C.~E. and {N{"o}the}, Maximilian and {Donath}, Axel and {Tollerud}, Erik and {Morris}, Brett M. and {Ginsburg}, Adam and {Vaher}, Eero and {Weaver}, Benjamin A. and {Tocknell}, James and {Jamieson}, William and {van Kerkwijk}, Marten H. and {Robitaille}, Thomas P. and {Merry}, Bruce and {Bachetti}, Matteo and {G{"u}nther}, H. Moritz and {Aldcroft}, Thomas L. and {Alvarado-Montes}, Jaime A. and {Archibald}, Anne M. and {B{'o}di}, Attila and {Bapat}, Shreyas and {Barentsen}, Geert and {Baz{'a}n}, Juanjo and {Biswas}, Manish and {Boquien}, M{'e}d{'e}ric and {Burke}, D.~J. and {Cara}, Daria and {Cara}, Mihai and {Conroy}, Kyle E. and {Conseil}, Simon and {Craig}, Matthew W. and {Cross}, Robert M. and {Cruz}, Kelle L. and {D'Eugenio}, Francesco and {Dencheva}, Nadia and {Devillepoix}, Hadrien A.~R. and {Dietrich}, J{"o}rg P. and {Eigenbrot}, Arthur Davis and {Erben}, Thomas and {Ferreira}, Leonardo and {Foreman-Mackey}, Daniel and {Fox}, Ryan and {Freij}, Nabil and {Garg}, Suyog and {Geda}, Robel and {Glattly}, Lauren and {Gondhalekar}, Yash and {Gordon}, Karl D. and {Grant}, David and {Greenfield}, Perry and {Groener}, Austen M. and {Guest}, Steve and {Gurovich}, Sebastian and {Handberg}, Rasmus and {Hart}, Akeem and {Hatfield-Dodds}, Zac and {Homeier}, Derek and {Hosseinzadeh}, Griffin and {Jenness}, Tim and {Jones}, Craig K. and {Joseph}, Prajwel and {Kalmbach}, J. Bryce and {Karamehmetoglu}, Emir and {Ka{l}uszy{'n}ski}, Miko{l}aj and {Kelley}, Michael S.~P. and {Kern}, Nicholas and {Kerzendorf}, Wolfgang E. and {Koch}, Eric W. and {Kulumani}, Shankar and {Lee}, Antony and {Ly}, Chun and {Ma}, Zhiyuan and {MacBride}, Conor and {Maljaars}, Jakob M. and {Muna}, Demitri and {Murphy}, N.~A. and {Norman}, Henrik and {O'Steen}, Richard and {Oman}, Kyle A. and {Pacifici}, Camilla and {Pascual}, Sergio and {Pascual-Granado}, J. and {Patil}, Rohit R. and {Perren}, Gabriel I. and {Pickering}, Timothy E. and {Rastogi}, Tanuj and {Roulston}, Benjamin R. and {Ryan}, Daniel F. and {Rykoff}, Eli S. and {Sabater}, Jose and {Sakurikar}, Parikshit and {Salgado}, Jes{'u}s and {Sanghi}, Aniket and {Saunders}, Nicholas and {Savchenko}, Volodymyr and {Schwardt}, Ludwig and {Seifert-Eckert}, Michael and {Shih}, Albert Y. and {Jain}, Anany Shrey and {Shukla}, Gyanendra and {Sick}, Jonathan and {Simpson}, Chris and {Singanamalla}, Sudheesh and {Singer}, Leo P. and {Singhal}, Jaladh and {Sinha}, Manodeep and {Sip{H{o}}cz}, Brigitta M. and {Spitler}, Lee R. and {Stansby}, David and {Streicher}, Ole and {{{S}}umak}, Jani and {Swinbank}, John D. and {Taranu}, Dan S. and {Tewary}, Nikita and {Tremblay}, Grant R. and {Val-Borro}, Miguel de and {Van Kooten}, Samuel J. and {Vasovi{'c}}, Zlatan and {Verma}, Shresth and {de Miranda Cardoso}, Jos{'e} Vin{'i}cius and {Williams}, Peter K.~G. and {Wilson}, Tom J. and {Winkel}, Benjamin and {Wood-Vasey}, W.~M. and {Xue}, Rui and {Yoachim}, Peter and {Zhang}, Chen and {Zonca}, Andrea and {Astropy Project Contributors}},
                title = "{The Astropy Project: Sustaining and Growing a Community-oriented Open-source Project and the Latest Major Release (v5.0) of the Core Package}",
              journal = {\apj},
                 year = 2022,
                month = aug,
               volume = {935},
               number = {2},
                  eid = {167},
                pages = {167},
                  doi = {10.3847/1538-4357/ac7c74},
        archivePrefix = {arXiv},
               eprint = {2206.14220},
         primaryClass = {astro-ph.IM},
               adsurl = {https://ui.adsabs.harvard.edu/abs/2022ApJ...935..167A}
        }

@ARTICLE{Hunter2007,
   author = {{Hunter}, J.~D.},
    title = "{Matplotlib: A 2D Graphics Environment}",
  journal = {Computing in Science and Engineering},
     year = 2007,
    month = may,
   volume = 9,
    pages = {90-95},
      doi = {10.1109/MCSE.2007.55},
   adsurl = {http://adsabs.harvard.edu/abs/2007CSE.....9...90H}
}

@article{Tange2011a,
  title = {GNU Parallel - The Command-Line Power Tool},
  author = {O. Tange},
  address = {Frederiksberg, Denmark},
  journal = {;login: The USENIX Magazine},
  month = {Feb},
  number = {1},
  volume = {36},
  url = {http://www.gnu.org/s/parallel},
  year = {2011},
  pages = {42-47},
  doi = {10.5281/zenodo.16303}
}

@article{metcalf19,
author = {Andrew J. Metcalf and Tyler Anderson and Chad F. Bender and Scott Blakeslee and Wesley Brand and David R. Carlson and William D. Cochran and Scott A. Diddams and Michael Endl and Connor Fredrick and Sam Halverson and Daniel D. Hickstein and Fred Hearty and Jeff Jennings and Shubham Kanodia and Kyle F. Kaplan and Eric Levi and Emily Lubar and Suvrath Mahadevan and Andrew Monson and Joe P. Ninan and Colin Nitroy and Steve Osterman and Scott B. Papp and Franklyn Quinlan and Larry Ramsey and Paul Robertson and Arpita Roy and Christian Schwab and Steinn Sigurdsson and Kartik Srinivasan and Gudmundur Stefansson and David A. Sterner and Ryan Terrien and Alex Wolszczan and Jason T. Wright and Gabriel Ycas},
journal = {Optica},
number = {2},
pages = {233--239},
publisher = {OSA},
title = {Stellar spectroscopy in the near-infrared with a laser frequency comb},
volume = {6},
month = {Feb},
year = {2019},
url = {http://www.osapublishing.org/optica/abstract.cfm?URI=optica-6-2-233},
doi = {10.1364/OPTICA.6.000233},
}

@ARTICLE{kanodia18,
   author = {{Kanodia}, S. and {Wright}, J.},
    title = "{Python Leap Second Management and Implementation of Precise Barycentric Correction (barycorrpy)}",
  journal = {Research Notes of the American Astronomical Society},
archivePrefix = "arXiv",
   eprint = {1801.01634},
 primaryClass = "astro-ph.EP",
     year = 2018,
    month = jan,
   volume = 2,
      eid = {4},
    pages = {4},
      doi = {10.3847/2515-5172/aaa4b7},
   adsurl = {https://ui.adsabs.harvard.edu/abs/2018RNAAS...2....4K}
}

@ARTICLE{turner16,
       author = {{Turner}, Jake D. and {Christie}, Duncan and {Arras}, Phil and
         {Johnson}, Robert E. and {Schmidt}, Carl},
        title = "{Investigation of the environment around close-in transiting exoplanets using CLOUDY}",
      journal = {\mnras},
         year = "2016",
        month = "Jun",
       volume = {458},
       number = {4},
        pages = {3880-3891},
          doi = {10.1093/mnras/stw556},
archivePrefix = {arXiv},
       eprint = {1603.01229},
 primaryClass = {astro-ph.EP},
       adsurl = {https://ui.adsabs.harvard.edu/abs/2016MNRAS.458.3880T}
}

@ARTICLE{stefansson16,
       author = {{Stefansson}, Gudmundur and {Hearty}, Frederick and {Robertson}, Paul and
         {Mahadevan}, Suvrath and {Anderson}, Tyler and {Levi}, Eric and
         {Bender}, Chad and {Nelson}, Matthew and {Monson}, Andrew and
         {Blank}, Basil and {Halverson}, Samuel and {Henderson}, Chuck and
         {Ramsey}, Lawrence and {Roy}, Arpita and {Schwab}, Christian and
         {Terrien}, Ryan},
        title = "{A Versatile Technique to Enable Sub-milli-Kelvin Instrument Stability for Precise Radial Velocity Measurements: Tests with the Habitable-zone Planet Finder}",
      journal = {\apj},
         year = 2016,
        month = dec,
       volume = {833},
       number = {2},
          eid = {175},
        pages = {175},
          doi = {10.3847/1538-4357/833/2/175},
archivePrefix = {arXiv},
       eprint = {1610.06216},
 primaryClass = {astro-ph.IM},
       adsurl = {https://ui.adsabs.harvard.edu/abs/2016ApJ...833..175S}
}

@article{stefansson20,
	doi = {10.3847/1538-3881/ab5f15},
	url = {https://doi.org/10.3847%2F1538-3881%2Fab5f15},
	year = 2020,
	month = {feb},
	publisher = {American Astronomical Society},
	volume = {159},
	number = {3},
	pages = {100},
	author = {Gudmundur Stefansson and Caleb Ca{\~{n}}as and John Wisniewski and Paul Robertson and Suvrath Mahadevan and Marissa Maney and Shubham Kanodia and Corey Beard and Chad F. Bender and Peter Brunt and J. Christopher Clemens and William Cochran and Scott A. Diddams and Michael Endl and Eric B. Ford and Connor Fredrick and Samuel Halverson and Fred Hearty and Leslie Hebb and Joseph Huehnerhoff and Jeff Jennings and Kyle Kaplan and Eric Levi and Emily Lubar and Andrew J. Metcalf and Andrew Monson and Brett Morris and Joe P. Ninan and Colin Nitroy and Lawrence Ramsey and Arpita Roy and Christian Schwab and Steinn Sigurdsson and Ryan Terrien and Jason T. Wright},
	title = {A Sub-Neptune-sized Planet Transiting the M2.5 Dwarf G 9-40: Validation with the Habitable-zone Planet Finder},
	journal = {The Astronomical Journal}
}

@ARTICLE{floriano19,
       author = {{Alonso-Floriano}, F.~J. and {Snellen}, I.~A.~G. and {Czesla}, S. and
         {Bauer}, F.~F. and {Salz}, M. and {Lamp{\'o}n}, M. and {Lara}, L.~M. and
         {Nagel}, E. and {L{\'o}pez-Puertas}, M. and {Nortmann}, L. and
         {S{\'a}nchez-L{\'o}pez}, A. and {Sanz-Forcada}, J. and
         {Caballero}, J.~A. and {Reiners}, A. and {Ribas}, I. and
         {Quirrenbach}, A. and {Amado}, P.~J. and {Aceituno}, J. and
         {Anglada-Escud{\'e}}, G. and {B{\'e}jar}, V.~J.~S. and
         {Brinkm{\"o}ller}, M. and {Hatzes}, A.~P. and {Henning}, Th. and
         {Kaminski}, A. and {K{\"u}rster}, M. and {Labarga}, F. and
         {Montes}, D. and {Pall{\'e}}, E. and {Schmitt}, J.~H.~M.~M. and
         {Zapatero Osorio}, M.~R.},
        title = "{He I {\ensuremath{\lambda}} 10 830 {\r{A}} in the transmission spectrum of HD209458 b}",
      journal = {\aap},
         year = 2019,
        month = sep,
       volume = {629},
          eid = {A110},
        pages = {A110},
          doi = {10.1051/0004-6361/201935979},
archivePrefix = {arXiv},
       eprint = {1907.13425},
 primaryClass = {astro-ph.EP},
       adsurl = {https://ui.adsabs.harvard.edu/abs/2019A&A...629A.110A}
}

@ARTICLE{Ninan20,
       author = {{Ninan}, Joe P. and {Stefansson}, Gudmundur and {Mahadevan}, Suvrath and {Bender}, Chad and {Robertson}, Paul and {Ramsey}, Lawrence and {Terrien}, Ryan and {Wright}, Jason and {Diddams}, Scott A. and {Kanodia}, Shubham and {Cochran}, William and {Endl}, Michael and {Ford}, Eric B. and {Fredrick}, Connor and {Halverson}, Samuel and {Hearty}, Fred and {Jennings}, Jeff and {Kaplan}, Kyle and {Lubar}, Emily and {Metcalf}, Andrew J. and {Monson}, Andrew and {Nitroy}, Colin and {Roy}, Arpita and {Schwab}, Christian},
        title = "{Evidence for He I 10830 {\r{A}} Absorption during the Transit of a Warm Neptune around the M-dwarf GJ 3470 with the Habitable-zone Planet Finder}",
      journal = {\apj},
         year = 2020,
        month = may,
       volume = {894},
       number = {2},
          eid = {97},
        pages = {97},
          doi = {10.3847/1538-4357/ab8559},
archivePrefix = {arXiv},
       eprint = {1910.02070},
 primaryClass = {astro-ph.EP},
       adsurl = {https://ui.adsabs.harvard.edu/abs/2020ApJ...894...97N}
}

@ARTICLE{Santiago24,
       author = {{Gully-Santiago}, Michael and {Morley}, Caroline V. and {Luna}, Jessica and {MacLeod}, Morgan and {Oklop{\v{c}}i{\'c}}, Antonija and {Ganesh}, Aishwarya and {Tran}, Quang H. and {Zhang}, Zhoujian and {Bowler}, Brendan P. and {Cochran}, William D. and {Krolikowski}, Daniel M. and {Mahadevan}, Suvrath and {Ninan}, Joe P. and {Stef{\'a}nsson}, Gu{\dj}mundur and {Vanderburg}, Andrew and {Zalesky}, Joseph A. and {Zeimann}, Gregory R.},
        title = "{A Large and Variable Leading Tail of Helium in a Hot Saturn Undergoing Runaway Inflation}",
      journal = {\aj},
         year = 2024,
        month = apr,
       volume = {167},
       number = {4},
          eid = {142},
        pages = {142},
          doi = {10.3847/1538-3881/ad1ee8},
archivePrefix = {arXiv},
       eprint = {2307.08959},
 primaryClass = {astro-ph.EP},
       adsurl = {https://ui.adsabs.harvard.edu/abs/2024AJ....167..142G}
}

@ARTICLE{Zhang23,
       author = {{Zhang}, Zhoujian and {Morley}, Caroline V. and {Gully-Santiago}, Michael and {MacLeod}, Morgan and {Oklop{\v{c}}i{\'c}}, Antonija and {Luna}, Jessica and {Tran}, Quang H. and {Ninan}, Joe P. and {Mahadevan}, Suvrath and {Krolikowski}, Daniel M. and {Cochran}, William D. and {Bowler}, Brendan P. and {Endl}, Michael and {Stef{\'a}nsson}, Gudmundur and {Tofflemire}, Benjamin M. and {Vanderburg}, Andrew and {Zeimann}, Gregory R.},
        title = "{Giant tidal tails of helium escaping the hot Jupiter HAT-P-32 b}",
      journal = {Science Advances},
         year = 2023,
        month = jun,
       volume = {9},
       number = {23},
          eid = {eadf8736},
        pages = {eadf8736},
          doi = {10.1126/sciadv.adf8736},
archivePrefix = {arXiv},
       eprint = {2306.03913},
 primaryClass = {astro-ph.EP},
       adsurl = {https://ui.adsabs.harvard.edu/abs/2023SciA....9F8736Z}
}

@ARTICLE{Palle20,
       author = {{Palle}, E. and {Nortmann}, L. and {Casasayas-Barris}, N. and {Lamp{\'o}n}, M. and {L{\'o}pez-Puertas}, M. and {Caballero}, J.~A. and {Sanz-Forcada}, J. and {Lara}, L.~M. and {Nagel}, E. and {Yan}, F. and {Alonso-Floriano}, F.~J. and {Amado}, P.~J. and {Chen}, G. and {Cifuentes}, C. and {Cort{\'e}s-Contreras}, M. and {Czesla}, S. and {Molaverdikhani}, K. and {Montes}, D. and {Passegger}, V.~M. and {Quirrenbach}, A. and {Reiners}, A. and {Ribas}, I. and {S{\'a}nchez-L{\'o}pez}, A. and {Schweitzer}, A. and {Stangret}, M. and {Zapatero Osorio}, M.~R. and {Zechmeister}, M.},
        title = "{A He I upper atmosphere around the warm Neptune GJ 3470 b}",
      journal = {\aap},
         year = 2020,
        month = jun,
       volume = {638},
          eid = {A61},
        pages = {A61},
          doi = {10.1051/0004-6361/202037719},
archivePrefix = {arXiv},
       eprint = {2004.12812},
 primaryClass = {astro-ph.EP},
       adsurl = {https://ui.adsabs.harvard.edu/abs/2020A&A...638A..61P}
}

@ARTICLE{Roberts23,
       author = {{Libby-Roberts}, Jessica E. and {Schutte}, Maria and {Hebb}, Leslie and {Kanodia}, Shubham and {Ca{\~n}as}, Caleb I. and {Stef{\'a}nsson}, Gu{\dh}mundur and {Lin}, Andrea S.~J. and {Mahadevan}, Suvrath and {Parts}, Winter and {Powers}, Luke and {Wisniewski}, John and {Bender}, Chad F. and {Cochran}, William D. and {Diddams}, Scott A. and {Everett}, Mark E. and {Gupta}, Arvind F. and {Halverson}, Samuel and {Kobulnicky}, Henry A. and {Kowalski}, Adam F. and {Larsen}, Alexander and {Monson}, Andrew and {Ninan}, Joe P. and {Parker}, Brock A. and {Ramsey}, Lawrence W. and {Robertson}, Paul and {Schwab}, Christian and {Swaby}, Tera N. and {Terrien}, Ryan C.},
        title = "{An In-depth Look at TOI-3884b: A Super-Neptune Transiting an M4Dwarf with Persistent Starspot Crossings}",
      journal = {\aj},
         year = 2023,
        month = jun,
       volume = {165},
       number = {6},
          eid = {249},
        pages = {249},
          doi = {10.3847/1538-3881/accc2f},
       adsurl = {https://ui.adsabs.harvard.edu/abs/2023AJ....165..249L}
}

@ARTICLE{Chakraborty25,
       author = {{Chakraborty}, H. and {Almenara}, J.~M. and {Lendl}, M. and {Ehrenreich}, D. and {Bouchy}, F. and {Bonfils}, X. and {Dancikova}, R. and {Deline}, A. and {Khan}, S. and {Netzel}, H. and {Shinde}, M. and {Verdier}, A.},
        title = "{The changing transit shape of TOI-3884 b}",
      journal = {\aap},
         year = 2025,
        month = oct,
       volume = {702},
          eid = {A100},
        pages = {A100},
          doi = {10.1051/0004-6361/202555998},
archivePrefix = {arXiv},
       eprint = {2509.04297},
 primaryClass = {astro-ph.EP},
       adsurl = {https://ui.adsabs.harvard.edu/abs/2025A&A...702A.100C}
}

@ARTICLE{Tamburo25,
       author = {{Tamburo}, Patrick and {Yee}, Samuel W. and {Garc{\'\i}a-Mej{\'\i}a}, Juliana and {Charbonneau}, David and {Bieryla}, Allyson and {Collins}, Karen A. and {Shporer}, Avi},
        title = "{Spot-crossing Variations Confirm a Misaligned Orbit for a Planet Transiting an M Dwarf}",
      journal = {\aj},
         year = 2025,
        month = oct,
       volume = {170},
       number = {4},
          eid = {200},
        pages = {200},
          doi = {10.3847/1538-3881/adf72f},
archivePrefix = {arXiv},
       eprint = {2506.11998},
 primaryClass = {astro-ph.EP},
       adsurl = {https://ui.adsabs.harvard.edu/abs/2025AJ....170..200T}
}

@ARTICLE{Mori25,
       author = {{Mori}, Mayuko and {Fukui}, Akihiko and {Hirano}, Teruyuki and {Narita}, Norio and {Livingston}, John H. and {Barkaoui}, Khalid and {Collins}, Karen A. and {de Leon}, Jerome P. and {Ikuta}, Kai and {Kawai}, Yugo and {Schwarz}, Richard P. and {Shporer}, Avi and {Srdoc}, Gregor},
        title = "{Multiband, Multiepoch Photometry of the Spot-crossing System TOI-3884: Refined System Geometry and Spot Properties}",
      journal = {\aj},
         year = 2025,
        month = oct,
       volume = {170},
       number = {4},
          eid = {204},
        pages = {204},
          doi = {10.3847/1538-3881/ade2df},
archivePrefix = {arXiv},
       eprint = {2506.06445},
 primaryClass = {astro-ph.EP},
       adsurl = {https://ui.adsabs.harvard.edu/abs/2025AJ....170..204M}
}

@ARTICLE{Almenara22,
       author = {{Almenara}, J.~M. and {Bonfils}, X. and {Forveille}, T. and {Astudillo-Defru}, N. and {Ciardi}, D.~R. and {Schwarz}, R.~P. and {Collins}, K.~A. and {Cointepas}, M. and {Lund}, M.~B. and {Bouchy}, F. and {Charbonneau}, D. and {D{\'\i}az}, R.~F. and {Delfosse}, X. and {Kidwell}, R.~C. and {Kunimoto}, M. and {Latham}, D.~W. and {Lissauer}, J.~J. and {Murgas}, F. and {Ricker}, G. and {Seager}, S. and {Vezie}, M. and {Watanabe}, D.},
        title = "{TOI-3884 b: A rare 6-R$_{E}$ planet that transits a low-mass star with a giant and likely polar spot}",
      journal = {\aap},
         year = 2022,
        month = nov,
       volume = {667},
          eid = {L11},
        pages = {L11},
          doi = {10.1051/0004-6361/202244791},
archivePrefix = {arXiv},
       eprint = {2210.10909},
 primaryClass = {astro-ph.EP},
       adsurl = {https://ui.adsabs.harvard.edu/abs/2022A&A...667L..11A}
}

@ARTICLE{Zechmeister14,
       author = {{Zechmeister}, M. and {Anglada-Escud{\'e}}, G. and {Reiners}, A.},
        title = "{Flat-relative optimal extraction. A quick and efficient algorithm for stabilised spectrographs}",
      journal = {\aap},
         year = 2014,
        month = jan,
       volume = {561},
          eid = {A59},
        pages = {A59},
          doi = {10.1051/0004-6361/201322746},
archivePrefix = {arXiv},
       eprint = {1311.5263},
 primaryClass = {astro-ph.IM},
       adsurl = {https://ui.adsabs.harvard.edu/abs/2014A&A...561A..59Z}
}

@ARTICLE{Nail25,
       author = {{Nail}, F. and {MacLeod}, M. and {Oklop{\v{c}}i{\'c}}, A. and {Gully-Santiago}, M. and {Morley}, C.~V. and {Zhang}, Z.},
        title = "{Cold dayside winds shape large leading streams in evaporating exoplanet atmospheres}",
      journal = {\aap},
         year = 2025,
        month = mar,
       volume = {695},
          eid = {A186},
        pages = {A186},
          doi = {10.1051/0004-6361/202452740},
archivePrefix = {arXiv},
       eprint = {2410.19381},
 primaryClass = {astro-ph.EP},
       adsurl = {https://ui.adsabs.harvard.edu/abs/2025A&A...695A.186N}
}

@ARTICLE{MacLeod25,
       author = {{MacLeod}, Morgan and {Oklop{\v{c}}i{\'c}}, Antonija and {Nail}, Fabienne and {Linssen}, Dion},
        title = "{Streams and Bubbles: Tidal Shaping of Hydrodynamic Planetary Outflows}",
      journal = {\apj},
         year = 2025,
        month = jul,
       volume = {988},
       number = {1},
          eid = {63},
        pages = {63},
          doi = {10.3847/1538-4357/ade0b7},
       adsurl = {https://ui.adsabs.harvard.edu/abs/2025ApJ...988...63M}
}

@ARTICLE{Getman25,
       author = {{Getman}, Konstantin V. and {Kochukhov}, Oleg and {Ninan}, Joe P. and {Feigelson}, Eric D. and {Airapetian}, Vladimir S. and {Waggoner}, Abygail R. and {Cleeves}, L. Ilsedore and {Forbrich}, Jan and {Dzib}, Sergio A. and {Law}, Charles J. and {Rab}, Christian and {Krolikowski}, Daniel M.},
        title = "{Multi-observatory Study of Young Stellar Energetic Flares (MORYSEF): No Evidence for Abnormally Strong Stellar Magnetic Fields after Powerful X-Ray Flares}",
      journal = {\apj},
         year = 2025,
        month = feb,
       volume = {980},
       number = {1},
          eid = {57},
        pages = {57},
          doi = {10.3847/1538-4357/ada55c},
archivePrefix = {arXiv},
       eprint = {2501.01321},
 primaryClass = {astro-ph.SR},
       adsurl = {https://ui.adsabs.harvard.edu/abs/2025ApJ...980...57G}
}

@INPROCEEDINGS{Shefov64,
       author = {{Shefov}, N.~N.},
        title = "{Helium in the upper atmosphere}",
    booktitle = {Theoretical Interpretation of Upper Atmosphere Emission},
         year = 1964,
       editor = {{Bates}, David Robert},
       series = {IAU Symposium},
       volume = {18},
        month = jan,
        pages = {73},
       adsurl = {https://ui.adsabs.harvard.edu/abs/1964IAUS...18...73S}
}

@ARTICLE{Adams11,
       author = {{Adams}, Fred C.},
        title = "{Magnetically Controlled Outflows from Hot Jupiters}",
      journal = {\apj},
         year = 2011,
        month = mar,
       volume = {730},
       number = {1},
          eid = {27},
        pages = {27},
          doi = {10.1088/0004-637X/730/1/27},
archivePrefix = {arXiv},
       eprint = {1101.4234},
 primaryClass = {astro-ph.EP},
       adsurl = {https://ui.adsabs.harvard.edu/abs/2011ApJ...730...27A}
}

@ARTICLE{Strugarek25arXiv250213262S,
       author = {{Strugarek}, A. and {Shkolnik}, E.},
        title = "{Introduction to magnetic star-planet interactions}",
      journal = {arXiv e-prints},
         year = 2025,
        month = feb,
          eid = {arXiv:2502.13262},
        pages = {arXiv:2502.13262},
          doi = {10.48550/arXiv.2502.13262},
archivePrefix = {arXiv},
       eprint = {2502.13262},
 primaryClass = {astro-ph.EP},
       adsurl = {https://ui.adsabs.harvard.edu/abs/2025arXiv250213262S}
}

\bibliographystyle{aasjournal}



\end{document}